\documentclass[10pt,journal,compsoc]{IEEEtran}
\ifCLASSOPTIONcompsoc
  \usepackage[nocompress]{cite}
\else
  \usepackage{cite}
\fi
\ifCLASSINFOpdf
   \usepackage[pdftex]{graphicx}
\else
\fi
\usepackage[T1]{fontenc}
\usepackage{cite}
\usepackage{blindtext}
\usepackage {tikz}
\usepackage{latexsym}
\usepackage{cite}
\usepackage{amsmath}
\usepackage{float}
\usepackage{pifont}
\usepackage{enumitem}
\usepackage{graphicx}
\usepackage{epstopdf}
\usepackage{amsthm}
\usepackage{caption}
\usepackage{subcaption}
\usepackage{ amssymb }
\usepackage{csquotes}
\usepackage[labelformat=simple]{subcaption}

\usepackage{array}

\usepackage{pdfsync}

\begin{document}
%
\title{Anchor Free IP Mobility}

\author{Mohammed Al-Khalidi, Nikolaos Thomos, Martin J. Reed, Mays F. AL-Naday and Dirk Trossen
	\IEEEcompsocitemizethanks{\IEEEcompsocthanksitem Mohammed Al-Khalidi, Nikolaos Thomos, Martin J. Reed and Mays F. AL-Naday are with the School of Computer Science and Electronic Engineering,	University of Essex,
		Colchester, CO4 3SQ, UK.\protect\\
		E-mails: {(mshawk,nthomos,mjreed,mfhaln)@essex.ac.uk}
		\IEEEcompsocthanksitem Dirk Trossen is with InterDigital Europe Ltd.,
		London, EC2A 3QR, UK. Email: Dirk.Trossen@InterDigital.com}
	\thanks{This work was carried out within the project POINT, which has received funding from the European Union's Horizon 2020 research and innovation programme under grant agreement No 643990.}}

\IEEEtitleabstractindextext{

\begin{abstract}
\boldmath
   Efficient mobility management techniques are critical in providing seamless connectivity and session continuity between a mobile node and the network during its movement.
   Current mobility management solutions generally require a central entity in the network core, tracking IP address movement and anchoring traffic from source to destination through point-to-point tunnels. Intuitively, this approach suffers from scalability limitations as it creates bottlenecks in the network, due to sub-optimal routing via the anchor point.  Meanwhile, alternative anchorless, solutions are not feasible due to the current limitations of the IP semantics, which strongly ties addressing information to location. 
   In contrast, novel path-based forwarding solutions may be exploited  for feasible anchorless solutions. In this paper, we propose a novel network-based mobility management solution that facilitates IP mobility over such a path-based forwarding substrate. Our solution exploits the advantages of such substrates in decoupling path calculation from data transfer to eliminate the need for anchoring traffic through the network core; thereby, allowing flexible path calculation and service provisioning. 
   Furthermore, by eliminating the limitation of routing via the anchor point, our approach reduces the network cost compared to anchored solution through bandwidth saving while maintaining comparable handover delay. We evaluate our solution through analytical and simulation models and compare it with the IETF standardized solution, Proxy Mobile IPv6 (PMIPv6). Evaluation results illustrate a significant saving in the total network cost when using our proposed solution, compared to its counterpart.

\end{abstract}


}

\maketitle


%
\IEEEpeerreviewmaketitle

\IEEEraisesectionheading{\section{Introduction}\label{sec:introduction}}

%
%
%
%

\IEEEPARstart{T}{he} significant progress achieved in mobile technologies, allowing users to enjoy Internet based content services during movement, relies on mobility management protocols. Mobility management is a challenging research topic since it largely affects users' experience in respect of preventing frequent disconnections and ensuring session continuity \cite{lee2010cost}. A key fact of the current Internet is that it was built on an architecture that exploits end host IP addresses as both communication endpoints and forwarding entities. This has been a fundamental obstruction in supporting many of the features and services
that came after its initial design, including end user mobility \cite{xylomenos2012caching}. Considering the rising volume of mobile traffic, due to increased content streaming, it can be concluded that the challenge  of supporting mobility will only grow bigger in the near future \cite{fotiou2013handling}. As predicted by Cisco, video traffic will compose 80 percent of all consumed Internet traffic in 2019 and traffic from wireless and mobile devices will rise to 66 percent of the total traffic \cite{index2015forecast}.

Conventional IP mobility techniques are based on functions existing in both the mobile terminal and the network to facilitate user mobility. Recently, due to the dominance of mobile traffic over the Internet, the new generation of wireless networks emphasize solutions that relocate mobility functions and procedures from the mobile device to network components. This approach, known as network-based mobility management, allows IP devices running standard protocol stacks to move freely between wireless access points belonging to the same local domain. Network-based mobility management is a desirable solution from a network operator's perspective because it allows service providers to enable mobility support without any user interaction or mobile node (MN) modification \cite{bernardos2010network} \cite{soto2010pmipv6}. For this purpose, several standardization bodies such as the Internet Engineering Task Force (IETF) and Third Generation Partnership Project (3GPP) are expending efforts on establishing reliable and efficient network-based mobility management services and protocols. However, many challenges still remain to be solved for achieving such a goal \cite{kong2008mobility}. 
 
 Proxy Mobile IPv6 (PMIPv6) \cite{gundavelli2008b} is the only IETF standardized network-based mobility management protocol until today and is aimed at accommodating various access technologies such as WiMAX, 3GPP, 3GPP2 and WLAN. In PMIPv6, a central Local Mobility Anchor (LMA) is responsible for maintaining reachability to the Mobile Node's (MN's) IP address while the MN moves between Mobile Access Gateways (MAGs) in the PMIPv6 domain by updating the binding cache in a binding table and maintaining a tunnel to the MAG for packet delivery. On the other hand, the MAG is responsible for detecting the MN's movement and initiating binding registration on behalf of the MN \cite{lee2013comparative} \cite{giust2014analytic}. Proxy Mobile IPv6 also supports IPv4 stack and dual stack mobility modes \cite{wakikawa2010ipv4}. PMIPv6, as with other IP mobility solutions, clearly increases network complexity. First of all, it violates network end-to-end transparency: although it provides user experience transparency, an essential goal for mobility support, it does not provide network addressing transparency which requires unaltered mechanisms for the flow of packets and unaltered logical addressing between source and destination \cite{carpenter2000internet}. PMIPv6 also increases network fragility due to the explosive growth of the binding table size in the LMA for all MNs in the domain. In addition, it imposes processing complexity in the network core (LMA) and edges (MAGs) to support the necessary protocol functionality during mobility \cite{alderson2005understanding} \cite{doyle2005robust}. Similar procedures are adapted by 3GPP in cellular networks where the mobility management entity (MME) controls mobility signaling on the control plane and the serving gateway (S-GW) anchors user traffic on the data plane using the General Packet Radio Service (GPRS) Tunnelling Protocol (GTP) \cite{lucent2009lte} to support mobility. GTP is a group of IP-based communications protocols used in the Global System for Mobile Communications (GSM), Universal Mobile Telecommunications System (UMTS) and Long-Term Evolution (LTE) core networks. In 3GPP architectures, GTP and Proxy Mobile IPv6 based interfaces are specified on various interface points \cite{ali2009network}.
 
The aforementioned mobility management approaches are mainly used to reduce mobility signaling costs in environments with a high mobility rate, but as a consequence they cause extra packet tunnelling overhead and inefficient routing due to central traffic anchoring in the network. Such drawbacks of current IP mobility solutions motivate the search for better approaches, as investigated in this paper. One promising approach involves utilizing  new forwarding architectures that rely purely on path information for the end-to-end forwarding of packets Instead of relying on host address-based communication with routing information distributed over various network elements. Solutions such as LIPSIN \cite{jokela2009lipsin}~\cite{reed2015stateless} and BIER~\cite{ietf-bier-problem-statement-00} utilize path information stored in the forwarded packet to deliver a packet traversing through the network.  In these alternative, path based approaches, the route computation determines an end-to-end path that is encoded into the packet header while the forwarding operation is considerably simpler than IP forwarding by virtue of executing a simple set membership test which can be efficiently implemented. Recent advances have shown how this path based approach can be carried out in commercially available SDN switches with a switching table size that is constant and considerably lower than traditional end-host address-based solutions~\cite{reed2015stateless}. Mobility in these architectures results in (partial) recomputation of a path with the opportunity to deliver the data over an optimal path after every handover operation.
 
 The main purpose of this paper is to propose and investigate a path-based approach to mobility management. However, to support the path-based forwarding, as with existing mobile IP solutions, a control plane is required. The investigation of a new control plane is out of the scope of this paper, thus, we base our proposal on an existing solution; namely Information Centric Networking (ICN)~\cite{xylomenos2014survey} \cite{PURSUITproject}, specifically that developed in PURSUIT \cite{PURSUITproject}. PURSUIT employs a Publish-Subscribe paradigm for a path-based information dissemination that names information at the network layer decoupling request resolution from data transfer in both time and space. The asynchronous nature of the Publish/Subscribe architecture simplifies resynchronization after MN handoffs and greatly facilitates mobility. However, clean-slate ICN architecture proposals such as PURSUIT have one significant drawback in that the network stack in every MN and server, together with application network interface code, have to be replaced. Therefore, IP-over-ICN~\cite{trossen2015ip} has emerged as a solution that aims at enabling individual operators to enhance their services by deploying a gateway-based architecture; this offers improved IP-based services with an ICN infrastructure at its heart without incurring any changes to the end-user equipment that use existing IP protocol stacks and connectivity. The combination of the opportunities arising from the path-based forwarding, with its direct path possibilities, and the backward compatibility of the IP-over-ICN solution poses the question: can this new form of delivery architecture improve the performance of IP mobility?
 
In this paper, we answer the aforementioned question by virtue of proposing a novel network-based mobility management approach using an IP-over-ICN network where an efficient path-based forwarding solution provides the core of the network, while exposing backward compatible IP communication at the edges. In the proposed solution, no traffic anchoring is required to support mobility at the network core, and no MN equipment modification or user interaction is required at the network edges. To evaluate our proposal, we analyse the mobility costs in an IP-over-ICN network using random walks on connected graphs and derive the corresponding cost functions in terms of signaling, packet delivery and handover latency costs. We compare the mobility costs with those of the IETF standardized network-based mobility management protocol, Proxy Mobile IPv6 (PMIPv6). We also conduct a discrete event simulation of both schemes to compare the MN mobility performance and verify the theoretical analysis.

The rest of the paper is structured as follows. Section 2 provides an overview of the utilized IP-over-ICN network architecture that forms the mobility management solution proposed in Section 3. Section 4 gives an overview of the improvements offered by the proposed IP-over-ICN  mobility management that is formally modelled through a cost analysis in Section 5 for the evaluation of the proposal. Section 6 presents and discusses the modelling and simulations results, while a survey of related work is provided in Section 7. Finally the paper is concluded in Section 8. 
        
    \section{IP-Over-ICN Networks}
\begin{figure}[t]
		 	\centering
		 	\includegraphics[width=\columnwidth]{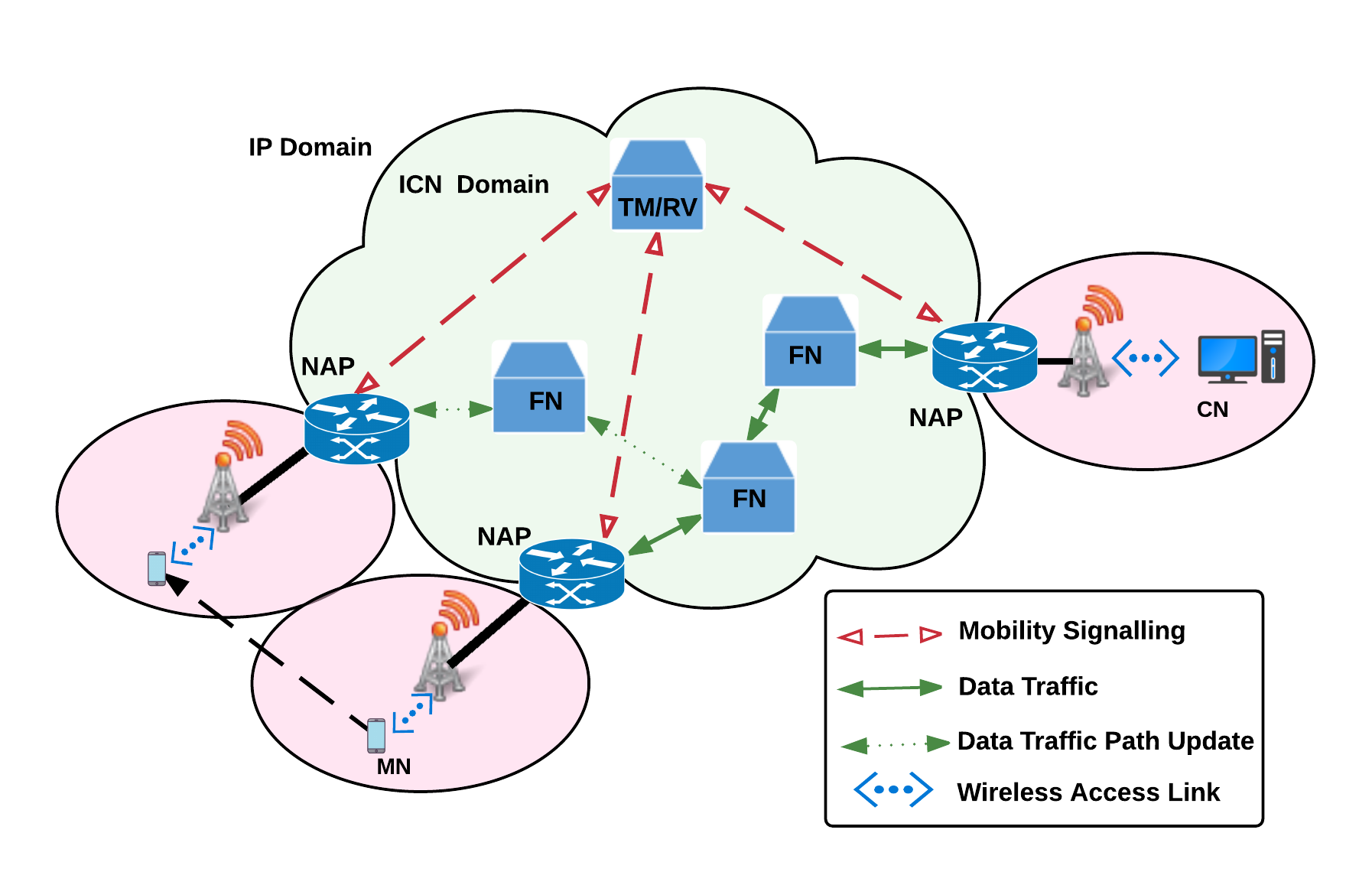}
		 	\caption{IP-over-ICN Architecture and Mobility Management Overview.}
		 	\label{IPoverICNArch}
		 	\vspace{-5mm}
		 	\vspace{-5mm}
		 \end{figure}

	As emphasized in the Introduction section, this paper uses a path-based approach to achieve its benefits. However, this path-based approach needs some form of architecture to manage the interface between IP forwarding and the path-based forwarding. An ICN architecture is utilized here as existing work has developed a suitable control-plane for the path-based forwarding. The proposed IP-over-ICN architecture follows a gateway-based approach, where the first link from the user device to the network uses existing IP-based protocols, such as HTTP, CoAP, TCP or IPv4/v6, while the Network Attachment Point (NAP) serves as an entry point to the ICN network and maps the chosen protocol abstraction to ICN. The ICN core employs a Publish-Subscribe paradigm \cite{trossen2012designing} for information dissemination that names information at the network layer, arranging individual information items into a context named \emph{scoping}. Scopes allow information items to be grouped according to application requirements, for example different categories of information. Relationships between information items and scopes are represented as a directed acyclic graph of which leaves represent pieces of information and inner nodes represent scopes. Each node in the graph is identified with its full path starting from a root scope, a more detailed explanation is given in \cite{trossen2012designing}. There are three main functional entities that compose the ICN architecture as shown in Fig.  \ref{IPoverICNArch}: the Rendezvous (RV), the Topology Manager (TM) and Forwarding Nodes (FN). The RV is responsible for matching publications and subscriptions of information items while the TM is responsible for constructing a delivery tree for the information object. This delivery tree is encoded in a forwarding identifier (FID) which is sent to the publisher so that it can forward the packets containing the information object to the subscriber. Note that the FID encodes a tree to allow for possible multicast delivery, where unicast is a trivial subset of a tree. In this paper we will ignore the multicast capability as we wish to compare with existing mobile IP solutions that are usually focused on unicast. In the network, there are also Forwarding Nodes (FN) that simply forward the information object to the subscriber using the specific FID generated for this transmission~\cite{reed2012traffic}. Throughout this paper, the TM and RV functions are assumed to be residing in the same entity for the sake of simplicity, although they may be distributed or separated to support a scalable and resilient solution.

The IP-over-ICN operation uses publish/subscribe (pub/sub) semantics for carrying IPv4/v6 datagrams over the ICN network. First, a na\"ive pub/sub signaling description will be given, to show the underlying principle, although in a likely deployment there will be optimizations to this na\"ive signaling that will be explained later. In the first instance, ICN signaling may sound complex. However, it must be remembered that this needs to be compared to the protocols required for an IP network application including DHCP, DNS, routing \emph{etc.} to name but a few for general support and of course the Proxy MIPv6 signaling that is the specific protocol relevant to this paper. ICN signaling may be likened to this support signaling.

\subsection{A na\"ive ICN signaling approach}
\label{sec:naive-icn-signalling}

To explain the underlying IP-over-ICN principle, the na\"ive signaling approach used is as described in \cite{trossen2015ip}. ICN uses a namespace to facilitate communication, this namespace may be used to represent any form of information. In an IP-over-ICN scenario, an IPv4/v6 address simply becomes an ICN name; the NAP uses publish/subscribe semantics to map IP datagrams to ICN names and then uses these names to forward IP datagrams as ICN information items through the ICN network. To aid the description, we will consider an IP client connected to what we describe as a client NAP (cNAP) and an IP server connected to a server NAP (sNAP). The cNAP and sNAP are only descriptive notations used for the na\"ive description, in practice a NAP will perform functions for any client or server connected to it so that, in practice, a NAP performs as both a cNAP and sNAP. An sNAP providing connectivity to an IP server is said to \emph{subscribe} to receive packets destined for the IP server, this subscription state is registered in the domain RV. Then if an IP client wishes to send data to the IP server the cNAP is said to \emph{publish} the IP datagrams to the IP server NAP. To actually forward the IP datagrams, the cNAP requires an FID for the forwarding function which is obtained through pub/sub matching. Pub/sub matching occurs in the RV when both a publisher and subscriber are registered for a unique ICN name, in this case the ICN name is the server IP address. Thus, when the cNAP registers the publication to the RV, the RV notes the match and requires the TM to send an appropriate FID to the cNAP so that it can publish (transmit) the data to the sNAP. In the na\"ive approach, when the IP server replies this whole mechanism can be reversed so that IP datagrams can flow in the reverse direction as well. When the client/server stop communicating (e.g. after a TCP FIN or after a suitable time-out) the publish/subscription matching state can be removed from the RV as communication is no longer required. The server subscription state is still maintained so that future IP clients can start a new communication.

In practice this na\"ive signaling approach is inefficient in terms of both state requirements in the RV and the number of signaling messages. Consequently, a practical system implements signaling optimizations including combining the cNAP publication message with an implicit subscription and only keeping the server subscription state in the RV. These optimizations are included in the signaling described in the following section.
\vspace{-5mm}

		 \begin{center}
		 	\begin{figure*}[t]
		 		\centering
		 		\includegraphics[width=0.8\textwidth,height=3.5in]{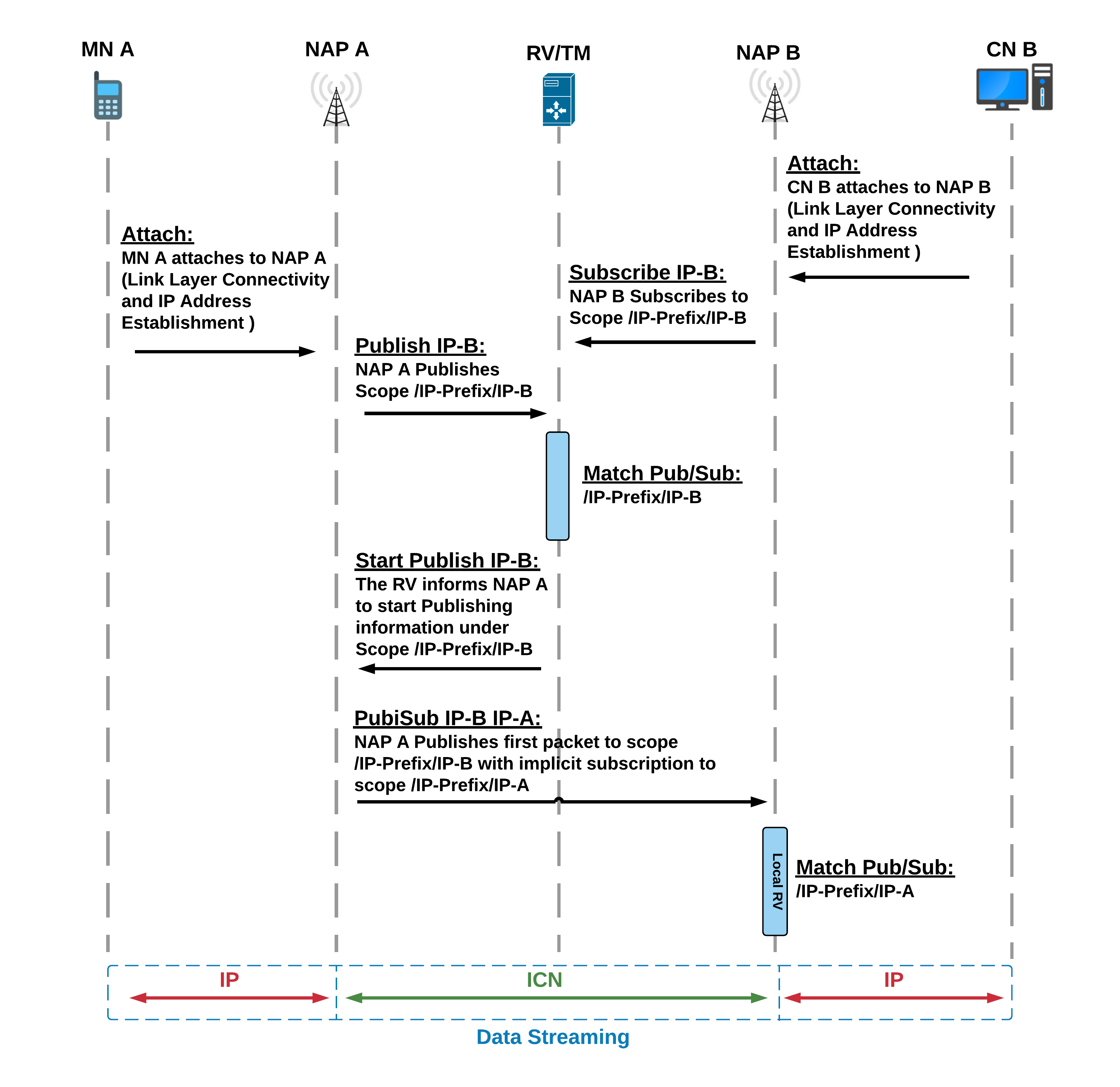}
		 		\caption{Sequence Diagram of Session Establishment in IP-over-ICN Networks.}
		 		\label{BBMR}
		 	\end{figure*}	
		 \end{center}
		 	
 	 	   			 \begin{center}
 	 	   			 	\begin{figure*}[t]
 	 	   			 		\centering
 	 	   			 		\includegraphics[width=0.8\textwidth,height=3.5in]{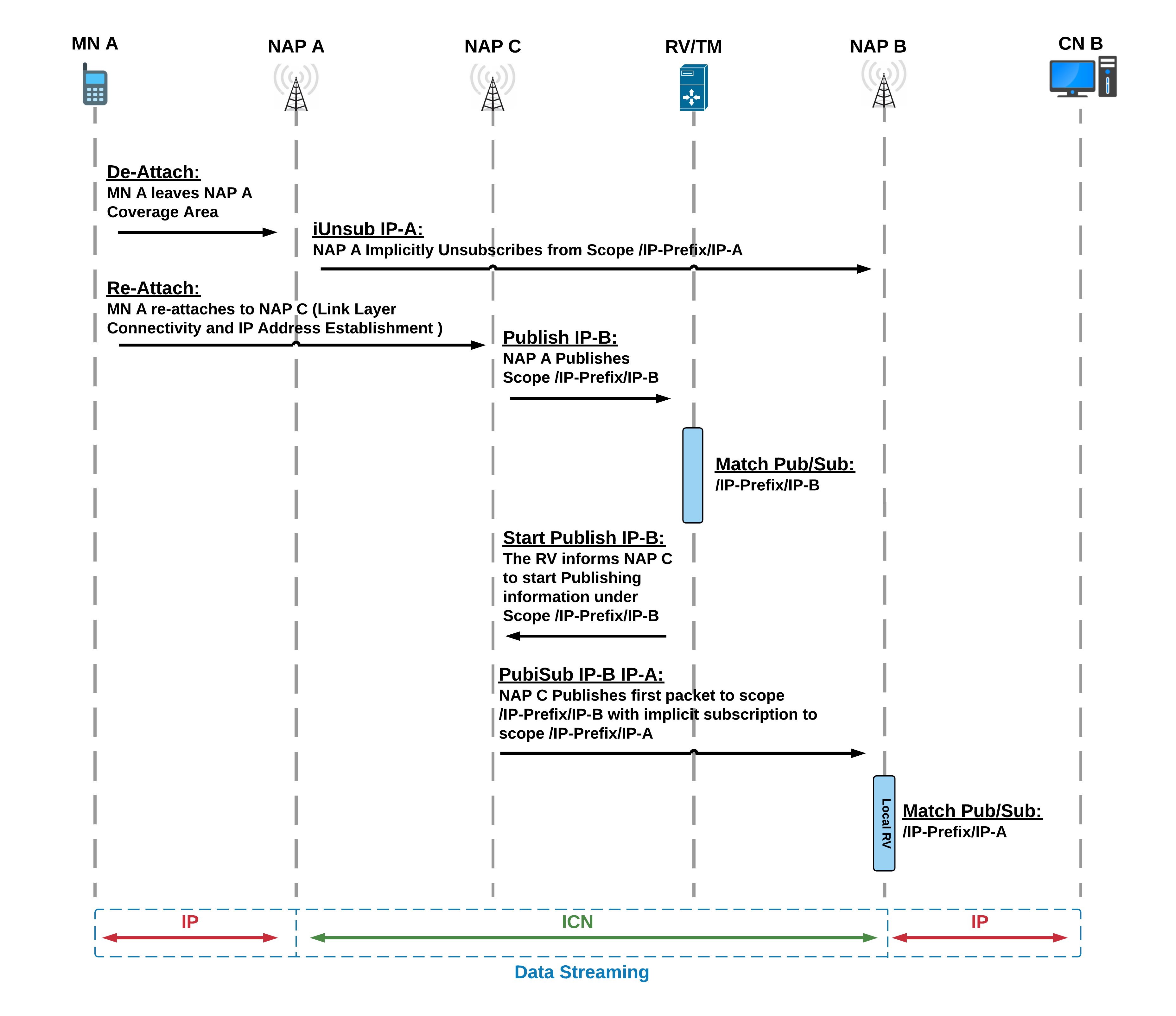}
 	 	   			 		\caption{Sequence Diagram of Handover Management in IP-over-ICN Networks.}
 	 	   			 		\label{BBMS}
 	 	   			 	\end{figure*}	
 	 	   			 \end{center} 
 	 \vspace{-5mm} 
 	 \vspace{-5mm} 			 
	 \section{Mobility Management in IP-over-ICN}
	 
\label{sec:mobil-manag-ip}	 

For mobility management in IP-over-ICN, we propose that the NAP could serve as a MAG that performs the mobility management on behalf of a mobile node. The NAP occupies a key role in both MN network attachment and IP/ICN abstraction and interfacing. Therefore, it is a natural point for detecting the mobile node's movements to and from the access link since it resides at the access link where the mobile node is attached. On the ICN side, we propose that a centralized TM initially sets up the required routing state in the network and creates FIDs to forward packets from a NAP to every other NAP according to the deployed routing algorithm. All the NAPs receive their specified FIDs and populate a local table containing the complete set of FIDs required to reach any other NAP in the network. In IP-over-ICN, the mobile node will receive the IPv4/IPv6 address that the NAP locally assigns, and the NAP will act on behalf of the mobile node as the publisher or the subscriber towards the ICN. The ICN represents the network structure of IP addresses in a namespace under a unique root scope and an IP address of any device is interpreted as an appropriate ICN name under this scope. This means that the NAP will be ready to receive any information being sent to the assigned IP address by determining the appropriate ICN name according to the defined namespace. Therefore, any IP packet being sent to an IP address allocated to an IP device will arrive at the NAP serving it as an ICN-compliant notification to a subscription to this IP address (represented as an appropriate ICN name) \cite{trossen2015ip}. The IP namespace proposed includes a network prefix scope identifier that serves as a root identifier and represents the IP network prefix allocated to serve the subject network domain. Under this root scope, there exists a so-called IP scope that represents the individual IP addresses allocated to IP endpoints that exist within the domain. These identifiers are formed by hashing a fully qualified IP address into a single 256 bit identifier.
	 
Fig. \ref{BBMR} shows a sequence diagram of the messages exchanged to establish a session between two IP endpoints in the proposed IP-over-ICN network. In this scenario, we assume that both the mobile node and the corresponding node are in the same network domain. For simplicity, the examples assume a single subnet where a MN is likely to keep its IP address when moving among NAPs. The ICN core maintains session continuity by maintaining the same pub/sub matching relations at the rendezvous even when a MN moves from one NAP to the other. This forms one of the IP-over-ICN advantages compared to Proxy MIPv6 networks for intra-domain scenarios because scalability is maintained by dividing and regionalizing the broadcast domain behind NAPs and routing is done through the ICN infrastructure using ICN semantics. This removes the scalability restrictions that would exist in an IP-core that would have to route /32 host-routes for every host in the domain. In the IP-over-ICN case, the external IP network could be divided into subnets (maybe for address allocation reasons). The IP-over-ICN will treat the IP addresses in the same manner as a single subnet as forwarding within the ICN is orthogonal to the IP address allocation.

For IP-over-ICN networks, end-node IP address sustainability can be maintained using any suitable IP auto-configuration mechanism suitable for the network infrastructure deployed. One example is the Dynamic Host Configuration Protocol (DHCP) where every NAP can act as a DHCP server serving the entire subnet deployed in the IP-over-ICN domain. We propose that every DHCP server can be configured to only assign local addresses (for MNs that locally attach to the NAP) from a specific pool within the subnet while it assigns addresses from outside the pool only to MNs that have previously been allocated an IP address at a previous NAP and intentionally ask for this specific IP address at the new NAP. This ensures that no IP address conflict would happen when the MN moves between NAPs. When a MN moves to a new NAP and goes into the DHCP RENEWING state, it would simply send a DHCPREQUEST message including the previously assigned IPv4 home address in the "Requested IP Address" option. The DHCPREQUEST is sent to the address specified in the Server Identifier option of the previously received DHCPOFFER and DHCPACK messages. The DHCP server would then send a DHCPACK to the MN to acknowledge the assignment of the committed IPv4 address following RFC2131 \cite{droms1997rfc} and RFC5844 \cite{wakikawa2010ipv4}. Each DHCP server on every NAP is configured to have the same IP address throughout the network, enabling the DHCPREQUEST message to be automatically sent to the available DHCP server on the access link without any delay. To facilitate IP address reuse, we propose that the Rendezvous keeps track of all IP addresses used to maintain pub/sub relations in the network and sends periodic reports to all DHCP servers notifying them of abandoned IP addresses. 

In the aforementioned scenario, first MN A attaches to NAP A (Link Layer Connectivity and IP Address Establishment). Then NAP A extracts from the first packet sent from MN A towards MN B the source and destination IP addresses. NAP A translates the extracted addresses into appropriate ICN names according to the defined IP namespace before publishing the destination address Scope /IP-Prefix/IP-B to the domain Rendezvous on behalf of MN A. Upon receiving this publication, the Rendezvous then matches it with a previous subscription of NAP B to the same scope on behalf of CN B. The Rendezvous triggers NAP A to start publishing information to the identified subscriber located at NAP B. NAP A then looks up it's local database for the appropriate FID to reach NAP B and uses it to send a PubiSub message directly to NAP B that includes the first data packet destined from MN A to CN B in addition to an implicit subscription to MN A's own scope /IP-Prefix/IP-A. NAP B utilizes it's local Rendezvous to maintain a match pub/sub relation for scope /IP-Prefix/IP-A, looks up it's local database for the appropriate FID to reach NAP A and uses this FID to start publishing information to the identified subscriber located at NAP A. At this point MN A and CN B can commence data exchange. This procedure is only required for the first data packet exchange between the two IP endpoints. Subsequent data packets can be directly sent using the allocated FIDs.

	 Fig. \ref{BBMS} shows a sequence diagram of the messages exchanged to manage a handover procedure for MN A from NAP A to NAP C. After initiating the handover procedure, the NAP on the previous link (NAP A) signals destination NAP B by sending an iUnsub message on behalf of MN A for it's own scope /IP-Prefix/IP-A. This way the local Rendezvous at NAP B can remove the subscription state for MN A. According to this example scenario, MN A re-attaches to NAP C and re-establishes Link Layer Connectivity and IP Address allocation through DHCP which triggers NAP C upon receiving the first IP packet from MN A to Publish the destination Scope /IP-Prefix/IP-B to the domain Rendezvous on behalf of MN A. The RV at this point re-matches the same publications and subscriptions established previously and triggers NAP C to start publishing information to the identified subscriber located at NAP B. NAP C then looks up its local database for the appropriate FID to reach NAP B and uses it to send a PubiSub message directly to NAP B that includes the first data packet destined from MN A to CN B in addition to an implicit subscription to MN A's own scope /IP-Prefix/IP-A. NAP B utilizes its local Rendezvous to maintain a match pub/sub relation for scope /IP-Prefix/IP-A, looks up its local database for the appropriate FID to reach NAP C and uses it to start publishing information to the identified subscriber located at NAP C. At this point MN A and CN B can commence data exchange without further disruption using MN A's new location. Fig. \ref{IPoverICNArch} shows the participating entities and communication message flows for each of the control and data planes during mobility.
			  
	 On the link layer, a number of metrics exist to indicate the quality of connection and are used to indicate mobility is occurring. One of these metrics is the Received Signal Strength Indicator (RSSI) which we use in this paper  -- alternatively, other predictors of mobility could be used, but their investigation is out of the scope of this paper. The RSSI value is part of the data transmitted by all mobile user equipment units. It is intended as a means to obtain a relative indication of the quality of connection that exists between the MN and the network access point it is connected to on the wireless network. This could be used as the trigger for movement described in this example. Which NAP a client connects to is almost entirely determined by the MN itself. Thus, when a client is given a choice between multiple NAPs offering the same service, it will always choose the NAP with the highest RSSI. On the other hand just like the initial association sequence, when a MN is moving it also uses RSSI to determine when to disassociate from a NAP and associate with another.
	 \vspace{-5mm} 
 \begin{center}
 \begin{figure}
 	
 	\begin{subfigure}[b]{0.45\textwidth}
 		\includegraphics[width=\textwidth]{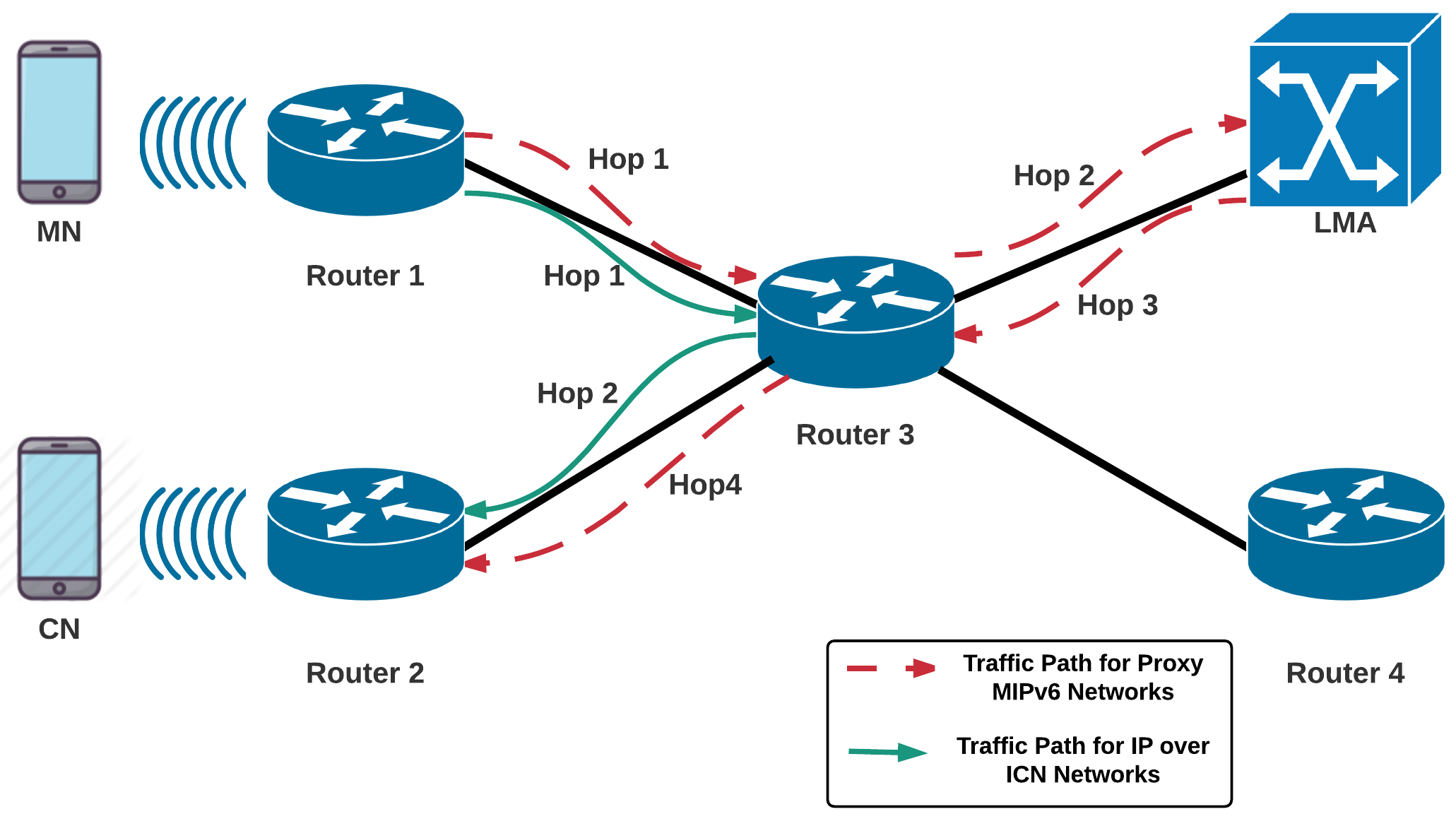}
 		\caption{Example Topology 1}
 		\label{fig:a123}
 	\end{subfigure}
 	
 	\begin{subfigure}[b]{0.45\textwidth}
 		\includegraphics[width=\textwidth]{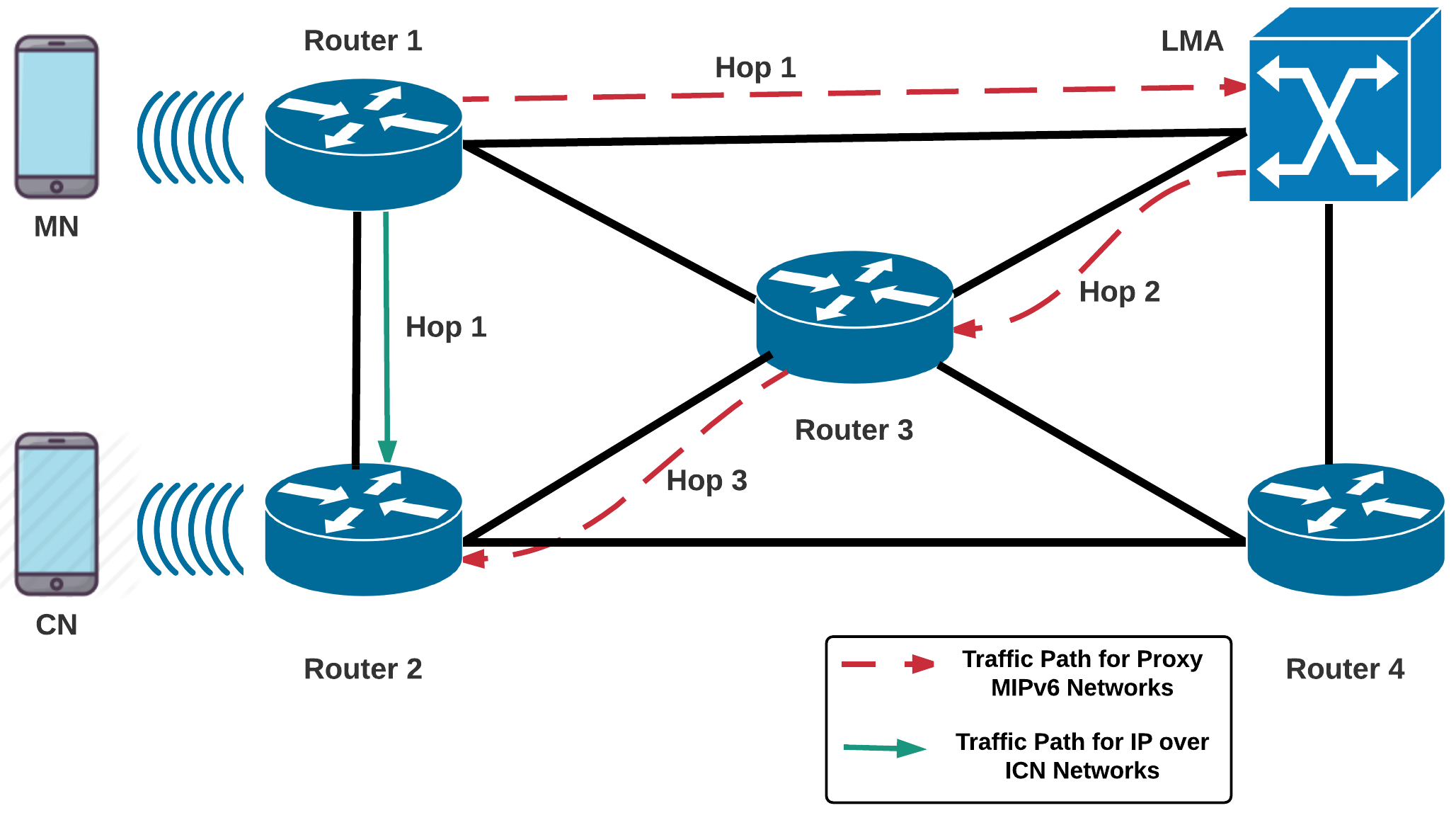}
 		\caption{Example Topology 2}
 		\label{fig:b123}		  
 	\end{subfigure}
 	\caption{Packet Delivery Routes in IP-over-ICN vs Proxy MIPv6 Networks.}
 	\label{fig:Motivation}
 	\vspace{-5mm}
 \end{figure}
\end{center}
\vspace{-5mm}	 
	 \section{Why IP-over-ICN for Network-Based Mobility Management? }
	 Proxy MIPv6 uses a centralized mobility management entity on both the data and control plane to facilitate network based mobility support. This approach on the one hand helps to reduce signaling costs in high mobility rate environments but on the other hand increases traffic and packet delivery cost within the networks core. Using this approach, all the traffic sent to and from a mobile node is driven through a local mobility anchor (LMA) that keeps track of the mobile nodes location and routes the traffic accordingly. This approach leads to using sub-optimal routes for packet delivery, thereby increasing the traffic overhead and end-to-end delay. The problem is evident, for example, when accessing a nearby server of a Content Delivery Network (CDN), or when receiving locally available IP multicast packets or sending IP multicast packets. A path-based approach on the other hand only requires a central point for mobility signaling (here IP-over-ICN) and delivery path creation, while the actual payload is delivered from source to destination through the shortest path without any anchoring.

	  To show the overhead caused by traffic anchoring in a simple way, we use the example in Fig. \ref{fig:Motivation}. As shown in the example, for a packet sent from a mobile node (MN) to reach a corresponding node (CN) in Fig. \ref{fig:a123} it crosses two routers (hops) in an IP-over-ICN network while it crosses four hops in Proxy MIPv6 networks to support network controlled mobility. Thus, the packet delivery cost using IP-over-ICN is half the cost of Proxy MIPv6 using this topology. The gain shown in this example is topology dependent as can be seen in Fig. \ref{fig:b123} where the number of hops crossed in an IP-over-ICN network is one hop versus three hops in Proxy MIPv6 networks. Therefore packet delivery cost in this IP-over-ICN scenario is one third the cost of the Proxy MIPv6 solution due to the fact that more links have been added to the same setup. An extended evaluation of network topology effect on router-level Internet performance has been shown in \cite{li2004first} and verified that many different graphs having the same distribution of node degree, may be considered opposites from the viewpoint of network engineering and result in widely varying end user bandwidths and router utilization distributions.
	 \vspace{-5mm}
	 \begin{figure}[t]
	 	\centering
	 	\includegraphics[width=3.5in]{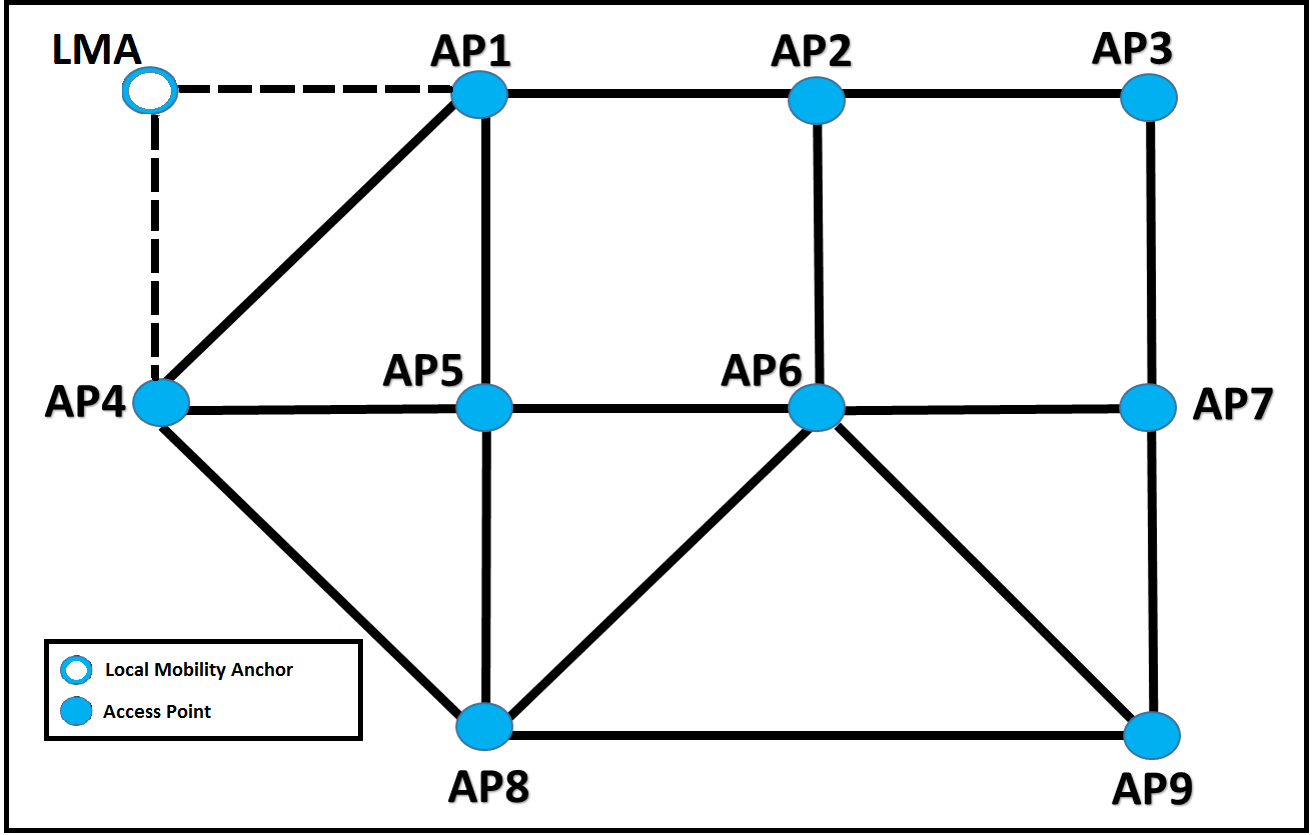}
	 	\caption{Network Model Example.}
	 	\label{top2}
	 	\vspace{-5mm}
	 \end{figure}

	 	 	 		\begin{center}
	 	 	 			\begin{table*}[ht]
	 	 	 				{\caption{Direction Probability Matrix and Steady State Probabilities.}
	 	 	 					\label{tab:2}
	 	 	 					\small
	 	 	 					\hfill{}
	 	 	 					\begin{tabular}{|c|c|c|c|c|c|c|c|c|c|c|}
	 	 	 						\hline
	 	 	 						Direction&AP1&AP2&AP3&AP4 & AP5 & AP6 & AP7 &AP8 & AP9 & Steady-State\\
	 	 	 						Probability $p_{(k,j)}$& & &  & & & &  & &  &                                         Probability $(\Pi)$\\
	 	 	 						\hline
	 	 	 						AP1& 1/4 &1/4 &0 & 1/4 &1/4 & 0 & 0 &0 & 0 & 0.100\\
	 	 	 						\hline
	 	 	 						AP2&1/4 &1/4 & 1/4 & 0 & 0 & 1/4 & 0 & 0 & 0 & 0.100\\
	 	 	 						\hline
	 	 	 						AP3&0 &1/3 & 1/3 & 0 & 0 & 0 & 1/3& 0 & 0 & 0.066 \\
	 	 	 						\hline
	 	 	 						AP4&1/4 &0 & 0 & 1/4 & 1/4 & 0 & 0 & 1/4& 0 & 0.100 \\
	 	 	 						\hline
	 	 	 						AP5&1/5 & 0 &0 &1/5 & 1/5 & 1/5 & 0 & 1/5& 0 & 0.133 \\
	 	 	 						\hline
	 	 	 						AP6&0&1/6 & 0 & 0 &1/6 & 1/6 & 1/6 & 1/6 & 1/6 & 0.166\\
	 	 	 						\hline
	 	 	 						AP7&0& 0 &1/4 &0 & 0 & 1/4 & 1/4 & 0 & 1/4 & 0.100 \\
	 	 	 						\hline
	 	 	 						AP8&0 &0 &0 & 1/5 &1/5 & 1/5 & 0 & 1/5& 1/5 & 0.133  \\
	 	 	 						\hline
	 	 	 						AP9&0 &0 &0 & 0 &0 & 1/4 & 1/4 & 1/4& 1/4 & 0.100  \\
	 	 	 						\hline
	 	 	 						\noalign{\smallskip}
	 	 	 						
	 	 	 					\end{tabular}}
	 	 	 					\hfill{}
	 	 	 				\end{table*}
	 	 	 			\end{center}	  	 
	 \vspace{-5mm}
	 \section{Mobility Modelling and Cost Analysis}
	 In order to analyze the mobility behavior of mobile nodes in Proxy MIPv6 and IP-over-ICN networks, a random walk mobility model is applied on connected graphs that represent the network topology in terms of wireless access points. This approach has been chosen due to the importance of the network topology and its influence on the total cost as described in the previous section. Fig. \ref{top2} shows an example network topology graph consisting of 10 nodes that will be used to explain the details of the analysis performed. Given a random starting point, we select a random neighbor to move into (assuming equal transition probability to any neighbor for simplicity), then we select a neighbor of this new point at random, and move to it \emph{etc.} The random sequence of points selected this way is a random walk on the graph. A random walk on a network graph of access points possesses some unique distinctive properties that can be pointed out, including that of spatial homogeneity. This means that the transition probability between two points ($x$ and $y$) on the graph should depend on their relative positions in space. This is obviously due to the fact that a mobile user can only move to a neighboring access point from any access point it is attached to at any given time. Also this implies that this random walk demonstrates the skip-free property, namely that to go from point $x$ to point $y$ it must pass through all intermediate points because it can only move one point at each step. In our analysis the wireless network is modelled as a connected graph whose nodes represent the coverage areas. This allows for flexibility in topology formation and cell shape assumptions from square and hexagonal cells to completely random topologies. Using a random walk on a connected graph to model user mobility leads to a discrete time finite Markov chain which provides a very practical and reliable way of estimating the location and direction probabilities of a moving user. The location probability represents the likelihood that a MN is located within the range of a specific access point at a given point in time, while the direction probability represents the likelihood that a MN is moving into the coverage area of a specific neighboring access point within the given set of neighboring access point at a given point in time. The Markov chain will be used to derive the global balance equations and also to introduce mobility rates into our mobility analysis. 

A random walk on a connected and undirected graph can be represented as follows \cite{choi2004combinatorial}. If $G = (V,E)$ is a connected, non-bipartite, undirected graph where $V$ are vertices that represent network access points and $E$ edges that represent the interconnections between the access points. Each access point, $k \in V$, has a set of neighbors $N_k=\left\{v : v\in V,\; (k,v)\in E\right\}$ with the number of neighbors denoted as $|N_k|$. This graph represents a Markov chain where the states are the nodes of $G$. Mobility is represented through elements $p_{(k,j)}$ of the direction probability matrix $\mathbf{P}=(p_{(k,j)}), \forall k,j \in V$, where $p_{(k,j)}$ is the probability that a MN was in the previous time slot within the range of an access point $k \in V$ and moves to an access point $j \in V$ in the current time slot. Given uniform probability of neighbor choice, $p_{(k,j)}$ depends on the degree, $|N_k|$, of a node $k$ by:
	 	\begin{equation}
	 	p_{(k,j)} = 
	 	\begin{cases}
	 	
	 	1/(|N_k|+1)  &\text{If $j \in N_k$}.\\
	 	\\
	 	
	 	0  &\text{Otherwise}.
	 	\end{cases}
	 	\end{equation}
Given the direction probability from the equation above, there exists a unique steady-state location probability distribution vector
	 		$\Pi$ = ($\pi_1$, $\pi_2$, . . . , $\pi_K$), such that $0 \le \pi_k \le 1$ for $1 \le k \le K$ and $\pi_k$ represents the probability of a node being located at $k \in V$.
	 		The steady-state probability vector can be obtained by solving
	 		$\Pi$ = $\Pi \mathbf{P}$ \cite{kleinrock1975queueing}. From our network model example in Fig. \ref{top2}, the direction probability matrix and the steady-state probability vector are shown in Table \ref{tab:2}. 

The mobility on the connected graph above can be represented as a Markov process where states represent the traversed network access points and transitions between states represent flows of a Markovian process. Fig.~\ref{fig:mc} shows a complete Markov chain representation of the example network topology shown in Fig. \ref{top2}. The Markov process introduces the mean cell border crossing rate $\mu$ where in the analysis we assume that the mean cell border crossing rate is the same in all cells; this essentially assumes that mobility is homogeneous and that cells have the same area. Note that the local mobility anchor (LMA) has not been included in the Markov chain as it is not part of the mobility model as no MN transition into the LMA is permitted.
	 	
	 	\usetikzlibrary {positioning}
	 	\definecolor {processblue}{cmyk}{0,0,0,0.50}
	 	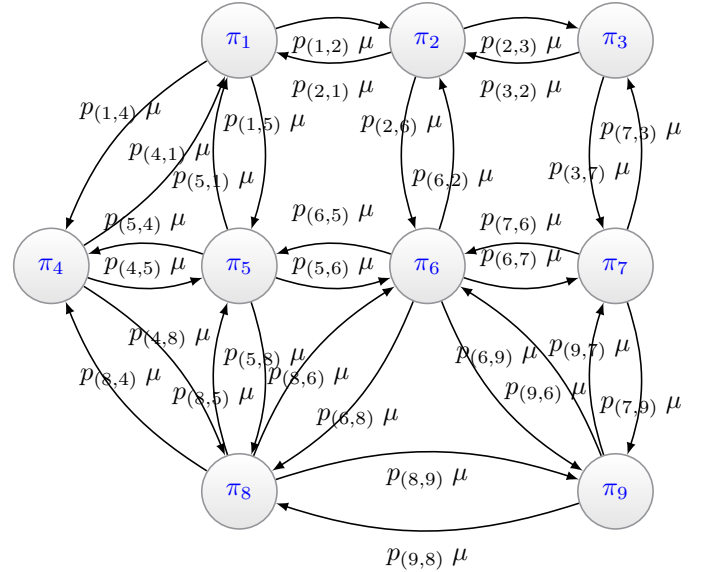
\begin{figure}
	 	\begin {center}
	 	
	 	\begin {tikzpicture}[-latex ,auto ,node distance =3 cm and 2.5cm ,on grid ,
	 	semithick ,
	 	state/.style ={ circle ,top color =white , bottom color = processblue!20 ,
	 		draw,processblue , text=blue , minimum width =1 cm}]
	 	\node[state] (C) {$\pi_5$};
	 	\node[state] (A) [above =of C] {$\pi_1$};
	 	\node[state] (B) [ right =of A] {$\pi_2$};
	 	\node[state] (I) [right =of B] {$\pi_3$};
	 	\node[state] (D) [below  =of B] {$\pi_6$};
	 	\node[state] (F) [left =of C] {$\pi_4$};
	 	\node[state] (G) [below  =of I] {$\pi_7$};
	 	\node[state] (H) [below  =of C] {$\pi_8$};
	 	\node[state] (E) [below  =of G] {$\pi_9$};

	 	\path (H) edge [bend right = -18] node[below =0.1 cm] {$p_{(8,9)}$ $\mu$} (E);
	 	\path (H) edge [bend left =18] node[below] {$p_{(8,6)}$ $\mu$} (D);
	 	\path (H) edge [bend left =18] node[below] {$p_{(8,5)}$ $\mu$} (C);
	 	\path (H) edge [bend left =18] node[above] {$p_{(8,4)}$ $\mu$} (F);
	 	
	 	\path (E) edge [bend right = -18] node[below =0.12 cm] {$p_{(9,8)}$ $\mu$} (H);
	 	\path (D) edge [bend left =18] node[below] {$p_{(6,8)}$ $\mu$} (H);
	 	\path (C) edge [bend left =18] node[above] {$p_{(5,8)}$ $\mu$} (H);
	 	\path (F) edge [bend left =18] node[above] {$p_{(4,8)}$ $\mu$} (H);
	 	
	 	\path (C) edge [bend right = 18] node[above =0.001 cm] { $p_{(5,4)}$ $\mu$} (F);
	 	\path (C) edge [bend right = -18] node[below =0.12 cm] {$p_{(5,1)}$ $\mu$} (A);
	 	\path (C) edge [bend right = 18] node[above =0.001 cm] {$p_{(5,6)}$ $\mu$} (D);
	 	
	 	\path (F) edge [bend right = 18] node[above =0.001 cm] { $p_{(4,5)}$ $\mu$} (C);
	 	\path (A) edge [bend right = -18] node[above =0.12 cm] {$p_{(1,5)}$ $\mu$} (C);
	 	\path (D) edge [bend right = 18] node[above =0.12 cm] {$p_{(6,5)}$ $\mu$} (C);
	 	
	 	\path (A) edge [bend right = 18] node[above =0.001 cm] { $p_{(1,4)}$ $\mu$} (F);
	 	\path (A) edge [bend right = -18] node[below =0.12 cm] {$p_{(1,2)}$ $\mu$} (B);
	 	
	 	\path (F) edge [bend right = 18] node[above =0.001 cm] { $p_{(4,1)}$ $\mu$} (A);
	 	\path (B) edge [bend right = -18] node[below =0.12 cm] {$p_{(2,1)}$ $\mu$} (A);
	 	
	 	\path (B) edge [bend right = 18] node[above =0.1 cm] { $p_{(2,6)}$ $\mu$} (D);
	 	\path (B) edge [bend right = -18] node[below =0.12 cm] {$p_{(2,3)}$ $\mu$} (I);
	 	
	 	\path (D) edge [bend right = 18] node[below =0.1 cm] { $p_{(6,2)}$ $\mu$} (B);
	 	\path (I) edge [bend right = -18] node[below =0.12 cm] {$p_{(3,2)}$ $\mu$} (B);

	 	\path (G) edge [bend right = 18] node[above =0.001 cm] { $p_{(7,3)}$ $\mu$} (I);
	 	\path (G) edge [bend right = -18] node[below =0.12 cm] {$p_{(7,9)}$ $\mu$} (E);
	 	\path (G) edge [bend right = 18] node[above =0.001 cm] {$p_{(7,6)}$ $\mu$} (D);
	 	
	 	\path (I) edge [bend right = 18] node[below =0.001 cm] { $p_{(3,7)}$ $\mu$} (G);
	 	\path (E) edge [bend right = -18] node[above =0.12 cm] {$p_{(9,7)}$ $\mu$} (G);
	 	\path (D) edge [bend right = 18] node[above =0.12 cm] {$p_{(6,7)}$ $\mu$} (G);
	 	
	 	\path (D) edge [bend right = 18] node[above =0.3 cm] { $p_{(6,9)}$ $\mu$} (E);
	 	\path (E) edge [bend right = 18] node[below =0.2 cm] { $p_{(9,6)}$ $\mu$} (D);	 	
	 \end{tikzpicture}
     \caption{Network Markov Chain Representation.}
	 \label{fig:mc}
     \vspace{-5mm}
	\end{center}
	\end{figure}
	
		\usetikzlibrary {positioning}
		\definecolor {processblue}{cmyk}{0,0,0,0.50}
		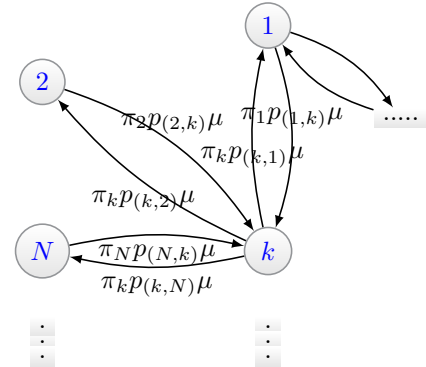
\begin{figure}[t]
			\begin {center}
			\begin {tikzpicture}[-latex ,auto ,node distance =3 cm and 3cm ,on grid ,
			semithick ,
			state/.style ={ circle ,top color =white , bottom color = processblue!20 ,
				draw,processblue , text=blue , minimum width =0 cm},MOH/.style={top color =white , bottom color = processblue!20 ,
				minimum width =0 cm}]
			\node[state] (C) {$N$};
			\node[state] (A) [left = 3 of B][above = 2.25 of C] {$2$};
			\node[state] (B) [above right =of C] {$1$};
			\node[state] (D) [right =of C] {$k$};
			
			\path (B) edge [bend left =18] node[above] {$\pi_{1} p_{(1,k)} \mu$} (D);
			\path (D) edge [bend right = -12] node[below =0.001 cm] {$\pi_k p_{(k,1)} \mu$} (B);
			\path (A) edge [bend left =18] node[above] {$\pi_{2} p_{(2,k)} \mu$} (D);
			\path (D) edge [bend right = -12] node[below =0.001 cm] {$\pi_k p_{(k,2)} \mu$} (A);
			\path (D) edge [bend right = -12] node[below =0.001 cm] {$\pi_k p_{(k,N)} \mu$} (C);
			\path (C) edge [bend right = -12] node[below =0.001 cm] {$\pi_N p_{(N,k)} \mu$} (D);

			\node[MOH] (H)  [above  right= 2.5 of D] {$.....$};
			\path (H) edge [bend left =18] node[above] {$ $} (B);
			\path (B) edge [bend left =18] node[above] {$ $} (H);
			
			\node[MOH] (E)  [below =1.2 of C] {$.$};
			\node[MOH] (EE)  [below =1 of C] {$.$};
			\node[MOH] (EEE)  [below =1.4 of C] {$.$};
			\node[MOH] (FF)  [below =1 of D]{$.$};
			\node[MOH] (FFf)  [below =1.2 of D]{$.$};
			\node[MOH] (FFff)  [below =1.4 of D]{$.$};
		\end{tikzpicture}
	\end{center}
	\caption{General Markov Process for Markov Chain Mobility.}
	 \label{fig:gmc}
	 \vspace{-5mm}
\end{figure}

    Assuming a system at steady state, the detailed balance equation for a user at state 1 (Network Access Point 1) can be obtained as follows:
\begin{multline}\label{emc2}
3 \pi_1 p_{(1,i)} = \sum\limits_{j \in J} \pi_j p_{(j,1)},\;\forall i \in J=\{2,4,5\}
\end{multline}
The general case for cell $k$, where the neighbors of the cell are $N_{k}$,  is represented in Fig.~\ref{fig:gmc}. Thus, the specific example in  (\ref{emc2}) can be generalized as the global balance equation:
\begin{equation}
	\vert{N_{k}}\vert \pi_k p_{k} \mu =  \sum_{j \in N_{k} } \pi_{j} p_{(j,k)} \mu \text{ ~for 0 $\le$ k$\le$ K}.
\end{equation}
where $p_{k} = p_{(k,1)} = p_{(k,2)} = \ldots =p_{(k,N_k)}$ is the generalized direction probability for a MN to move out of cell $k$.

\begin{table}[h]
	\caption{Summary of notation.}
	\label{tab:1}
	\centering
	\begin{tabular}{|c||c|}
		\hline
		Notation    &   Description\\
		\hline
		$\mu$  &   Mobility Rate\\
		\hline
		$p_{(k,j)}$ & Direction Probability that a mobile node is moving \\
		& into MAG $j$ from $k$\\
		\hline
		$\pi_{k}$ &  Location Probability that a mobile node is attached \\
		& to MAG k\\
		\hline
		$\Omega$, $\Omega^\prime$  &  Total Cost in a PMIPv6, IP-over-ICN \\
		& Network Respectively \\
		\hline
		$\Upsilon$, $\Upsilon^\prime$ &  Mobility signaling Cost in a PMIPv6,\\ 
		& IP-over-ICN Network Respectively\\
		\hline 	
		$\Lambda$, $\Lambda^{\prime}$ &  Mobility packet delivery cost in a PMIPv6, \\
		& IP-over-ICN Network Respectively. \\
		\hline
		$h_{k,a}$  &  Number of hops between the MN initial MAG $k$\\
		& and the LMA\\
		\hline
		$h_{j,a}$ & Number of hops between the MN new\\
		& MAG $j$ and the LMA \\
		\hline
		$h_{s,a}$  &  Number of hops between the CN's MAG\\
		& and the LMA\\
		\hline
		$R$, $R^{\prime}$ & Average packet rate in a PMPv6, \\
		& IP-over-ICN Network Respectively. \\
		\hline
		$O_{k}$, $O_{k}^{\prime}$ & Direct path packet overhead in a PMIPv6, \\
		& IP-over-ICN Network Respectively. \\
		\hline
		$h_{j,v}$ & Number of hops between NAP $j$\\
		& and the RV/TM\\ 
		\hline
		$h_{k,s}$ & Number of hops between NAP $k$\\
		& and the destination NAP $s$\\
		\hline
		$h_{j,s}$ & Number of hops between NAP $j$\\
		& and the destination NAP $s$\\
		\hline
		\noalign{\smallskip}
	\end{tabular}
\end{table}

 \begin{figure}[t]
 	\centering
 	\includegraphics[width=3.7in]{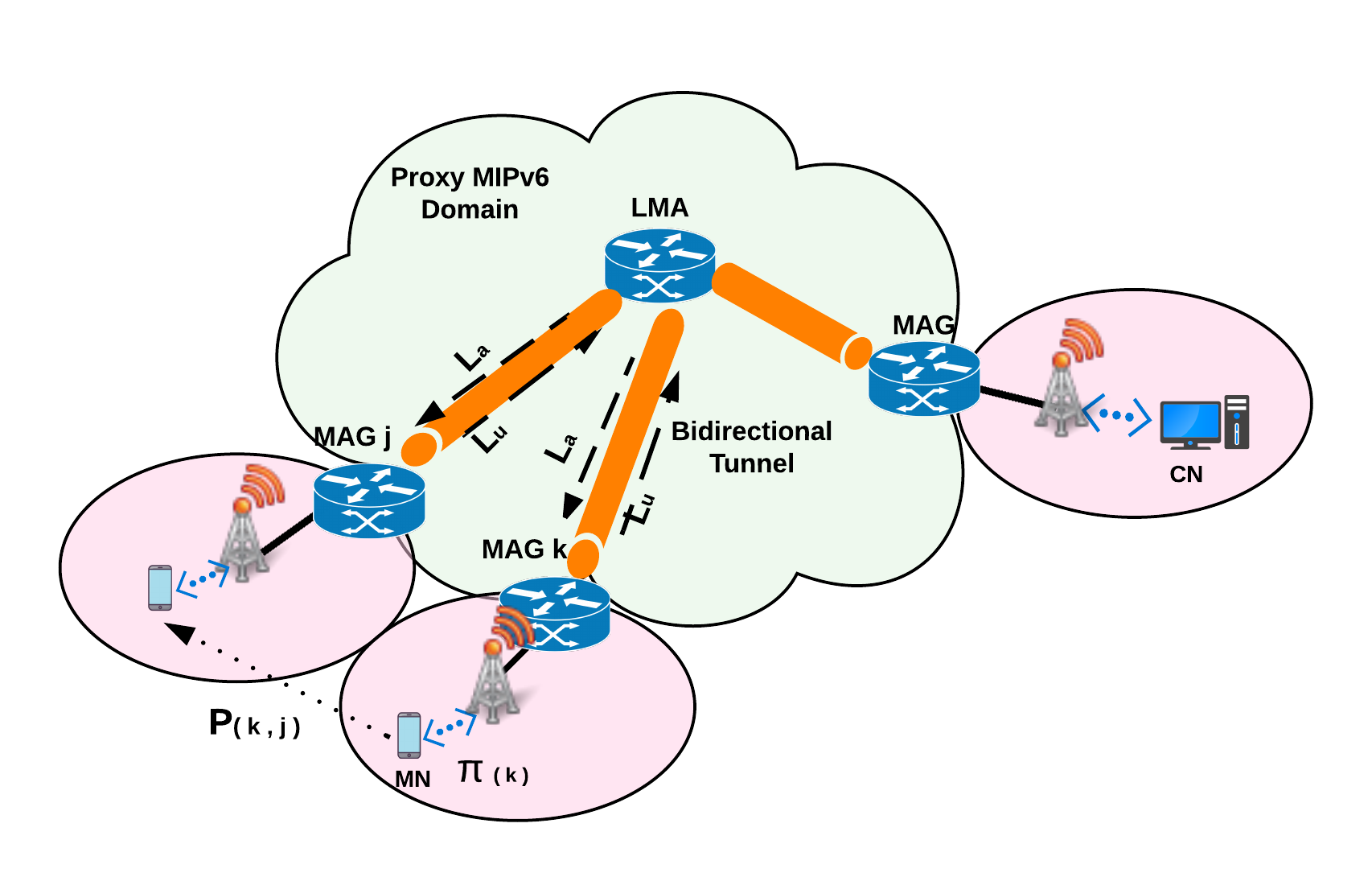}
 	\caption{Node Mobility in a Proxy MIPv6 Domain.}
 	\label{dat22}
 	\vspace{-5mm}
 \end{figure}

\subsection{Proxy MIPV6 Mobility Cost Analysis}
Proxy MIPv6 is used as a reference model to compare the performance of the proposed IP-over-ICN mobility solution. Proxy MIPv6 introduces two main entities that are involved in maintaining network enabled mobility support in a Proxy MIPv6 domain which are the LMA that represents the networks central mobility anchor point and the MAG that acts as a mobility proxy on behalf of the mobile node. In order to update the LMA about the MN's current location, a Proxy Binding Update (PBU) message is sent from the MAG to the MN's LMA. After accepting this PBU, the LMA sends back a Proxy Binding Acknowledgement (PBA) message to the MN's MAG that includes the MN's home network prefix. It also creates a binding cache entry into its binding cache table and establishes a bidirectional tunnel to the MAG. When the MN changes its point of attachment, the previous MAG on the previous access link detects the MN's detachment from the link and signals the LMA to remove the existing binding and routing state for that MN. The new MAG, upon detecting the MN on its access link, signals the LMA to update the binding state. Therefore, for every MN transition from one MAG to another, two mobility signaling events are required, one for each of the two participating MAG's to the domain's LMA \cite{yan2013localized}. Fig. \ref{dat22} shows a mobility scenario in a Proxy MIPv6 domain with one MN and a static corresponding node (CN). This scenario is considered in our mobility cost analysis. 

\begin{figure}[t]
\captionsetup[subfigure]{justification=justified,singlelinecheck=false}	
	\begin{subfigure}[b]{0.39\textwidth}
		\includegraphics[width=\textwidth]{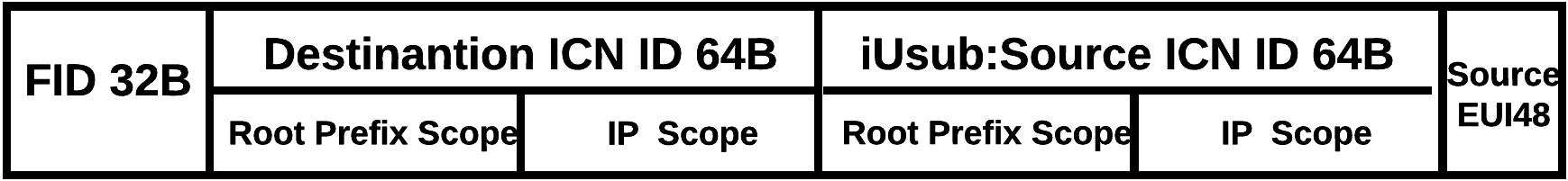}
		\caption{iUnsub Message ($\ell_{u}$)}
		\label{fig:a}
		\vspace*{2mm}
	\end{subfigure}
	
	\begin{subfigure}[b]{0.36\textwidth}
		\includegraphics[width=\textwidth]{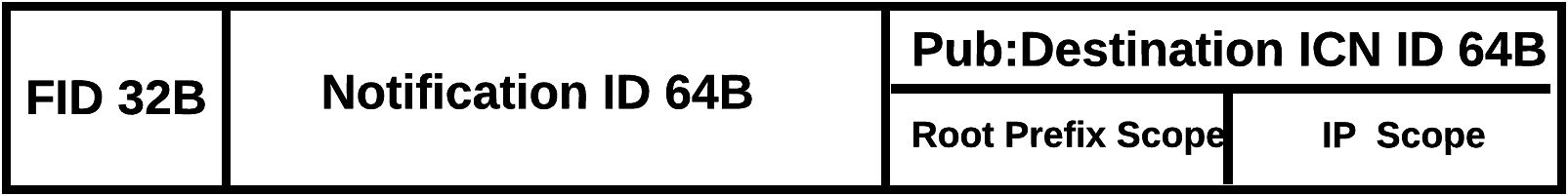}
		\caption{Request Publish Message ($\ell_{r}$)}
		\label{fig:b}
		\vspace*{2mm}
	\end{subfigure}
	
	\begin{subfigure}[b]{0.39\textwidth}
		\includegraphics[width=\textwidth]{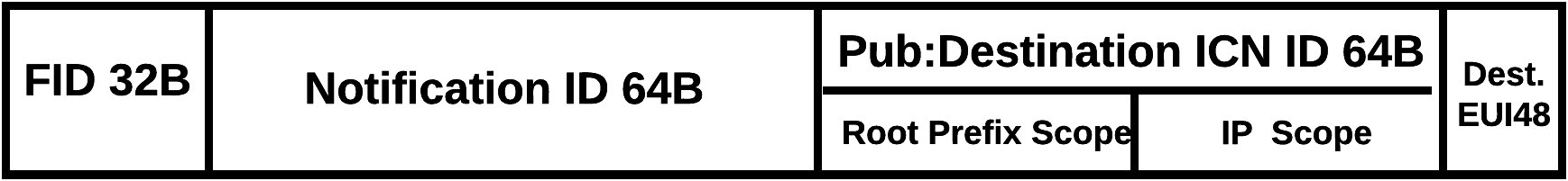}
		\caption{Start Publish Message ($\ell_{s}$)}
		\label{fig:c}
		\vspace*{2mm}
	\end{subfigure}
	
	\begin{subfigure}[b]{\columnwidth}
		\includegraphics[width=\textwidth]{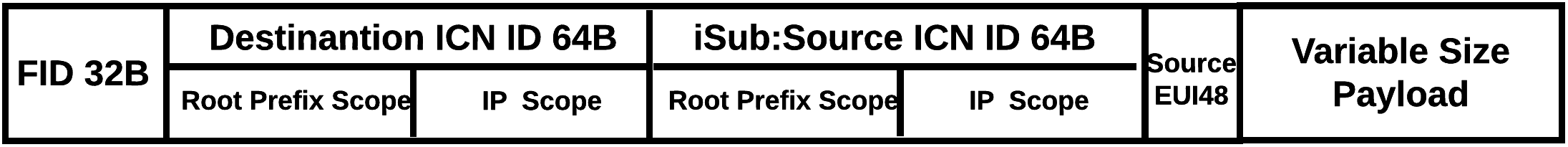}
		\caption{PubiSub Message ($\ell_{p}$)}
		\label{fig:d}
		\vspace*{2mm}
	\end{subfigure}	
	
	\begin{subfigure}[b]{0.3\textwidth}
		\includegraphics[width=\textwidth]{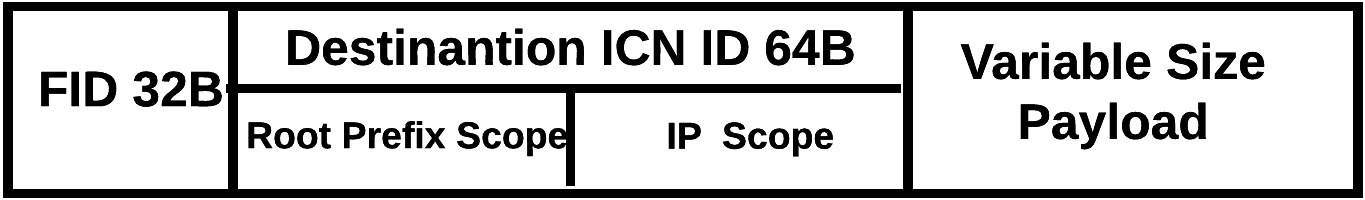}
		\caption{Data Payload Message ($\zeta$)}
		\label{fig:d}
		\vspace*{2mm}
	\end{subfigure}		
	
	\caption{IP-over-ICN Message Formats.}\label{fig:dat55}
\end{figure}

It can be concluded from Section 4 that no general closed form formula can be provided for mobility cost analysis, due to the high dependability of the total cost on the network topological aspects. Therefore, the mobility cost analysis will be conducted by feeding the topological factors into the cost analysis equations. Specifically, the total cost for PMIPv6 $\Omega$ is split into signaling $\Upsilon$ and packet delivery cost $\Lambda$ as follows. \footnote{Please refer to Table \ref{tab:1} for a a list of the notations used in this paper.}
\begin{equation}
	\Omega = \Upsilon + \Lambda
\end{equation}
where the signaling cost $\Upsilon$ is the signaling overhead for supporting mobility services for a MN. $\Lambda$ is the packet delivery cost that captures the tunneling and traffic anchoring overhead. $\Upsilon$ is calculated as the product of the size of mobility signaling messages and the hop distance. While $\Lambda$ is calculated as the product of the total packet size (including tunneling overhead) and the hop distance. The signaling cost $\Upsilon$ in Hops.Bytes$/$s can be calculated as:
\begin{multline}\label{emc6}
		\Upsilon = \sum_{k=1}^{K} \sum_{j \in N_{k}} \Big\{ \pi_{k} p_{(k,j)} \mu \: \big( h_{k,a} (|L_{u}| + |L_{a}|) \\
		+ h_{j,a} (|L_{u}| + |L_{a}|)\big) \Big\} 
\end{multline}
where $\pi_{k}$ is the location probability of a MN at MAG $k$; $p_{(k,j)}$ is the direction probability for the MN to move into MAG $j$ coverage area from MAG $k$; $\mu$ the MN's mobility rate of transition through a cell; $h_{k,a}$ is the number of hops between the LMA and MAG $k$; and, $h_{j,a}$ is the number of hops between the LMA and MAG $j$. As the MN changes its point of attachment, the previous MAG (i.e., MAG $k$) sends the de-registration PBU message to the LMA to inform the detachment of the MN at the access network managed by MAG $k$. As the new MAG (i.e., MAG $j$) detects the movement of the MN, it registers the MN to the LMA by sending a PBU message. $|L_{u}|$ is the size of the proxy binding update (PBU) message sent from the MAG to the LMA and $|L_{a}|$ is the size of the proxy binding acknowledgment (PBAck) message. \footnote{In this paper, $|x|$ denotes the length of message $x$.} A list of mobility messages and their corresponding sizes is shown in Table \ref{tab:3}. If we set the proxy binding update and proxy binding acknowledgment size (in bytes) as
\begin{equation}
\begin{split}
|L_{T}| = |L_{u}| + |L_{a}|
\end{split}
\end{equation}
and substitute with $|L_{T}|$ in (\ref{emc6}) we derive   
\begin{equation}\label{emc7}
	\begin{split}
		\Upsilon = \sum_{k=1}^{K} \sum_{j \in N_{k}} \pi_{k} p_{(k,j)} \mu |L_{T}|\: (h_{k,a} + h_{j,a}) 
	\end{split}
\end{equation}

The packet delivery cost $\Lambda$ is mainly used to investigate the tunneling and packet delivery overhead and is calculated as the product of total IPv6 packet size (including tunneling overhead) and the hop distance. The packet delivery cost for PMIPv6  $\Lambda$ is given by
\begin{equation}\label{emc8}
	\Lambda = \sum_{k=1}^{K} \pi_{k}\ R\ O_{k} 
\end{equation}
where R is the average packet rate, and O is the direct path packet cost in PMIPv6 which is obtained as
\vspace{-5mm}

\begin{equation}
\begin{split}	
O_{k} =	h_{s,a} ( \varphi + \zeta ) + h_{k,a} ( \varphi + \zeta )  
\end{split}	
\end{equation}
The parameter $h_{s,a}( \varphi + \zeta)$ is the direct path packet cost from the corresponding node (CN) to the LMA  and is equal to the number of hops between the CN and the LMA  $h_{s,a}$ multiplied by the average data packet length in Bytes including tunnelling overhead $( \varphi + \zeta)$. On the other hand, $h_{k,a}( \varphi + \zeta)$ denotes the direct path packet cost from the MN (k) to the LMA and therefore the cost is equal to the number of hops between the MN and the LMA, $h_{k,a}$, multiplied by the average data packet length in bytes including tunneling overhead $( \varphi + \zeta)$. The complete path packet cost is the sum of the cost between the CN and the LMA and the MN, $k$ and the LMA.
\begin{table}[t]
	\caption{List of Mobility Messages and their Sizes.}
	\label{tab:3}
	\centering
	\begin{tabular}{|c||c||c|}
		\hline
		Notation    &   Description   & Size in Bytes\\
		\hline
		$L_u$ & Proxy binding update (PBU) & 76 \cite{lee2010cost} \cite{lee2010much}\\
		\hline
		$L_a$ & Proxy binding acknowledgement & 76 \cite{lee2010cost} \cite{lee2010much}\\
		\hline
		$\zeta$ & Average payload length & 1024 \cite{lee2010cost} \cite{lee2010much}\\
		\hline 
		$\varphi$ & Proxy MIPv6 tunnelling header & 40 \cite{lee2010cost} \cite{lee2010much}\\
		\hline	
		$\ell_{u}$   & Implicit Unsubscribe (iUnsub) message& 166 \\
		\hline
		$\ell_{r}$   & Publish Request message & 160 \\
		\hline
		$\ell_{s}$   & Start Publish message & 166 \\
		\hline
		$\ell_{p}$   & Publish with Implicit Subscription & 166 \\
		& message (PubiSub)&  \\
		\hline
		$\varphi^{\prime}$   & ICN payload packet header & 96 \\
		\hline
		\noalign{\smallskip}
	\end{tabular}
	\vspace{-5mm}
\end{table}
\vspace{-5mm}
\begin{center}
	\begin{figure*}
		\begin{subfigure}[b]{0.48\textwidth}
			\includegraphics[width=\columnwidth]{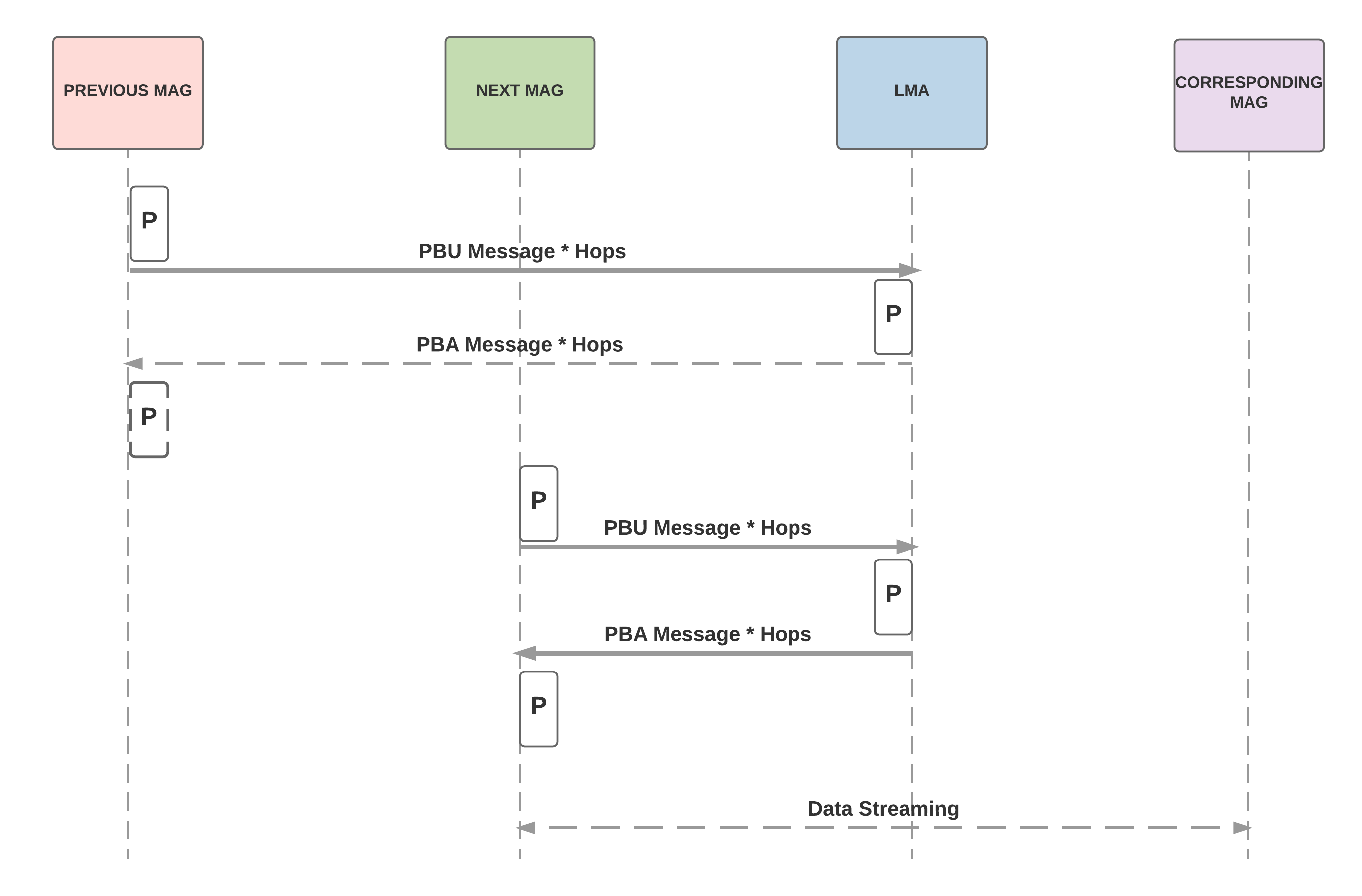}
			\caption{ Proxy MIPv6.}
			\label{light1}
		\end{subfigure}
		~
		\begin{subfigure}[b]{0.52\textwidth}
			\includegraphics[width=\columnwidth]{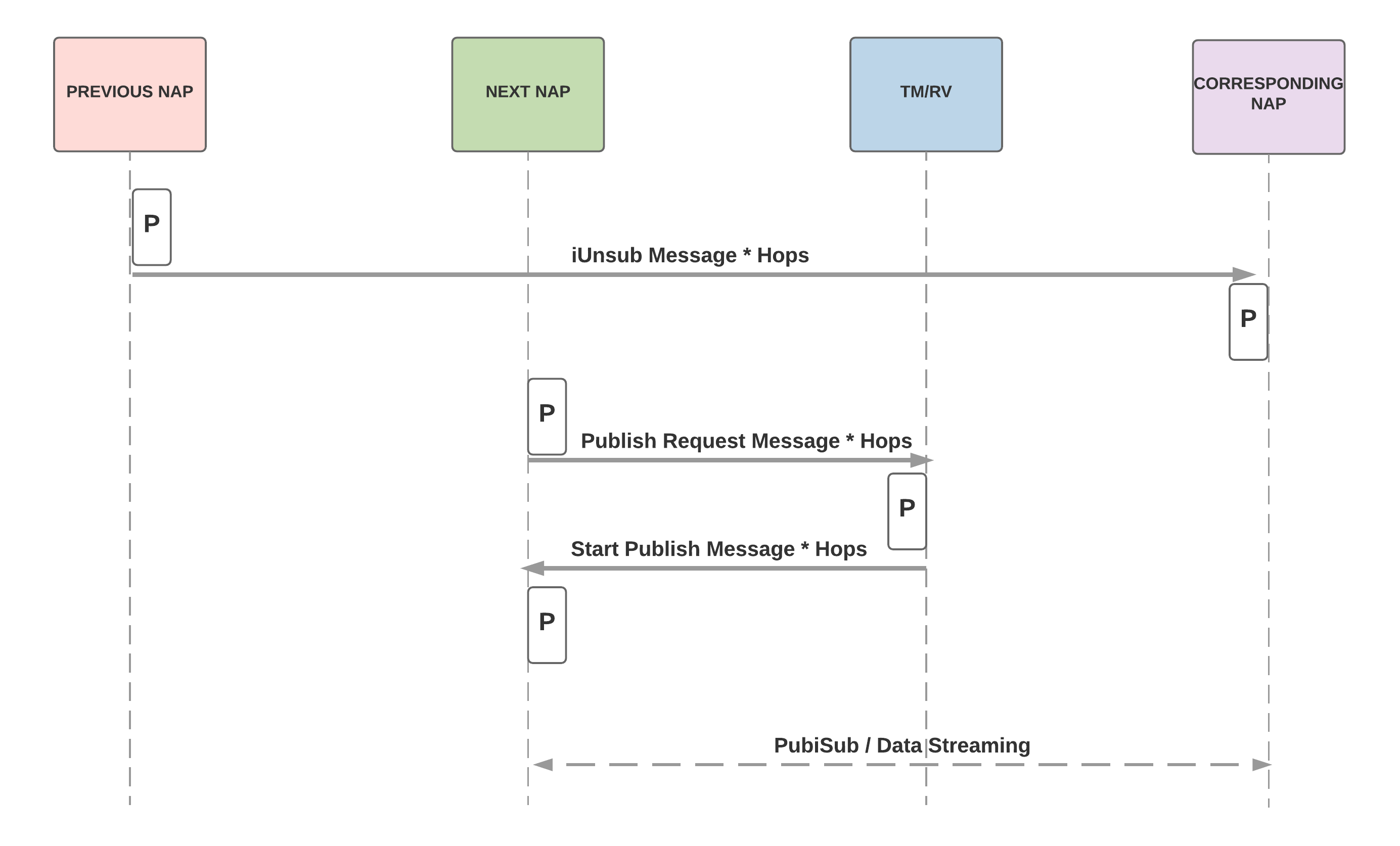}
			\caption{ IP over ICN. } 
			\label{light2}
		\end{subfigure}
		\caption{Handover Latency Analysis.}
		\label{light}
	\end{figure*}
\end{center}

\subsection{IP-Over-ICN Mobility Cost Analysis}
The mobility messages in the proposed IP-over-ICN infrastructure are totally incurred within the ICN core. Fig. \ref{BBMS} shows the sequence of mobility messages that take place during handover in an IP-over-ICN domain. For simplicity we always assume in our analysis that only one end of the communication is mobile (MN) and that the corresponding node (CN) is static and not generating any mobility signaling. After initiating a handover procedure, the NAP on the previous link (i.e. NAP A) signals the destination NAP (i.e. NAP B) by sending an iUnsub message ($\ell_{u}$) from the MN's own IP address scope. This enables NAP B to gracefully remove the subscription state for MN A from the CN's IP address scope. This state was established prior to the handover at NAP B's local RV. Upon the MN re-attachment to a new NAP (NAP C), and after it establishes Layer 2 connectivity and IP address allocation, NAP C receives the first IP packet destined to the CN and sends a Publish request message ($\ell_{r}$) to the domain RV requesting publication to the CN's IP address Scope. Upon receiving the publish request, the RV matches it with a previous subscription to the same address scope requested by NAP B and sends a start publish message ($\ell_{s}$) to the NAP on the new link (NAP C). NAP C then locally looks up the appropriate FID to reach NAP B and uses it to send a PubiSub message ($\ell_{p}$) to NAP B  that includes the first data packet from the MN to the CN in addition to an implicit subscription to MN A's own IP address scope. The PubiSub message triggers NAP B to utilize its local Rendezvous to maintain a match pub/sub relation for the mentioned scope, looks up its local database for the appropriate FID to reach NAP C and uses it to start publishing information to the identified subscriber. At this point MN A and CN B can commence sending and receiving data payload messages ($\zeta$). Fig. \ref{fig:dat55} illustrates the detailed message formats and sizes for the mobility messages needed in an IP-over-ICN setup.

The mobility signaling cost equals the size of the signaling messages multiplied by the number of hops. Therefore, the introduced signaling overhead is given by
	\begin{equation}\label{emc12}
		\begin{split}
			\Upsilon^\prime = \sum_{k=1}^{K} \sum_{j \in N_{k}} \Big\{ \pi_{k} p_{(k,j)} \mu \: \big(h_{k,s} |\ell_{u}| + h_{j,v} (|\ell_{r}| + |\ell_{s}|) \\
			 + h_{j,s} |\ell_{p}|\big) \Big\} 
		\end{split}
	\end{equation}
where $h_{k,s}$ is the number of hops between the previous NAP $k$ and the destination NAP s, $h_{j,v}$ is the number of hops between the new NAP $j$ and the RV/TM  and $h_{j,s}$ is the number of hops between NAP $j$ and the destination NAP $s$. $|\ell_{u}|$ denotes the size of an implicit unsubscribe (iUnsub) message sent from NAP $k$ to NAP $s$ when the MN initiates a handover. $|\ell_{r}|$ is the size of a publish request message sent from NAP $j$ to the RV/TM upon a change in the MN's NAP attachment requesting publication to the destination address scope. $|\ell_{s}|$ stands for the size of a start publish message sent from the domain RV/TM after a match pub/sub happens triggering NAP $j$ to start sending data packets to NAP $s$. Finally, $|\ell_{p}|$ is the size of a publish with implicit subscribe (PubiSub) message sent from NAP $j$ towards NAPs including the first data payload in addition to an implicit subscription to the MN's address scope at the new location (NAP $j$). In the upcoming evaluations the payload size has been excluded from the $\ell_{p}$ message size as it is not considered a mobility signaling cost.
	
The packet delivery cost, $\Lambda^{\prime}$, is mainly used to investigate the packet delivery overhead and is calculated as the product of total packet size and the hop distance:
	\begin{equation}\label{emc15}
		\Lambda^{\prime} = \sum_{k=1}^{K} \pi_{k}\ R^{\prime}\ O_{k}^{\prime}
	\end{equation}
where $R^{\prime}$ is the average packet rate in an IP-over-ICN network, and $O_{k}^{\prime}$ is the direct path packet overhead in IP-over-ICN obtained as
	\begin{equation}
		O_{k}^{\prime} = h_{s,k}( \varphi^{\prime} + \zeta )
	\end{equation}
where $h_{s,k}$ is the number of hops between NAP $s$ where the CN is attached and NAP $k$ where the MN is attached, and $\varphi^{\prime}$ is the size of the ICN packet header. Finally $\zeta$ is the average payload length in Bytes.

\begin{figure}[t]
	\centering
	\includegraphics[width=\columnwidth]{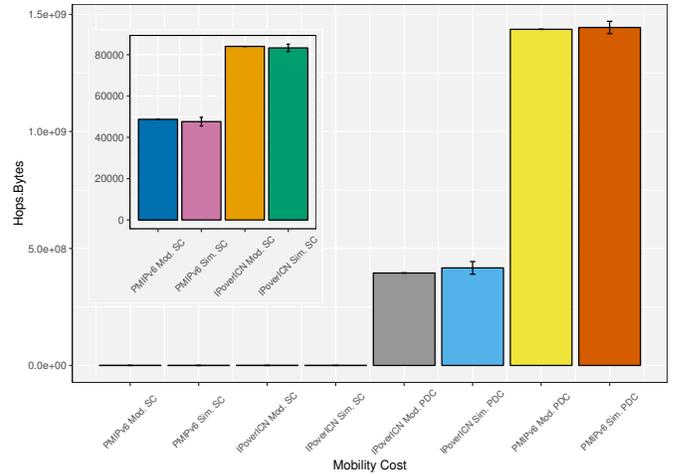}
	\caption{Modelling (Mod.) and simulation (Sim.) of a single MN at 70 miles/h with total packet delivery cost (PDC) and signaling cost (SC) using the example topology of section 5. The error bars show the 95\% confidence interval of the simulation results.}
	\label{fig:simvsmodel1}
\end{figure}

\vspace{-5mm}
\subsection{Handover Latency Analysis}
In this section, we analyze the latency differences between Proxy MIPv6 and IP-over-ICN. To allow a simple analysis, latency is interpreted in terms of number of messages exchanged, processes required and hops traversed to facilitate a successful mobility handover. According to the sequence diagrams in Fig. 10, we assume that $p$ denotes message processing time, $m$ denotes the time to exchange a message and $h$ denotes the number of hops that a message traverses. For simplicity we will assume that $p$ and $m$ are represented in arbitrary time units with $p = m = 1$ time unit i.e. that a link transmission delay is comparable to forwarding delay. Therefore, for Proxy MIPv6, the handover latency cost $T_{c}$ according to the message sequence in  Fig. \ref{light1} and the hop notations in Table \ref{tab:1} would be
\begin{equation}\label{emc13}
T_{c}= 5p + mh_{k,a} + 2mh_{j,a}
\end{equation}
The PBA message and its subsequent processing in dashed line (according to Fig. \ref{light1}) between the LMA and the previous MAG has not been added to the latency cost in (\ref{emc13}) as it does not impact the time consumed by the MN to re-establish the session on the new MAG. For IP-over-ICN, the latency cost $T_{c}^{\prime}$ according to the message sequence in  Fig. \ref{light2} and also referring to the hop definitions in Table \ref{tab:1} would be 
\begin{equation}\label{emc14}
T_{c}^{\prime}= 5p + mh_{k,s} + 2mh_{j,v}
\end{equation}
where the PubiSub message sent at the end of an IP over ICN handover has not been added to the latency cost in (\ref{emc14}) as it carries the MN's first data payload and therefore does not incur any extra latency.  Although, (13) and (14) have the same number of node processes, the costs $T_c$ and $T_c^{'}$ are not necessarily equal to each other, as the path lengths may not be equal. If the LMA (in PMIPv6) and the RV/TM (in IP-over-ICN) are the same location, then the last term may be the same in both cases, however, the middle term differs as $h_{k,a}$ represents the source to anchor hop-count, whereas $h_{k,s}$ represents the source to corresponding node hop-count.
\vspace{-5mm}

\begin{center}
	\begin{figure*}
		
		\begin{subfigure}[b]{0.5\textwidth}
			\includegraphics[width=\columnwidth,height=2.8in]{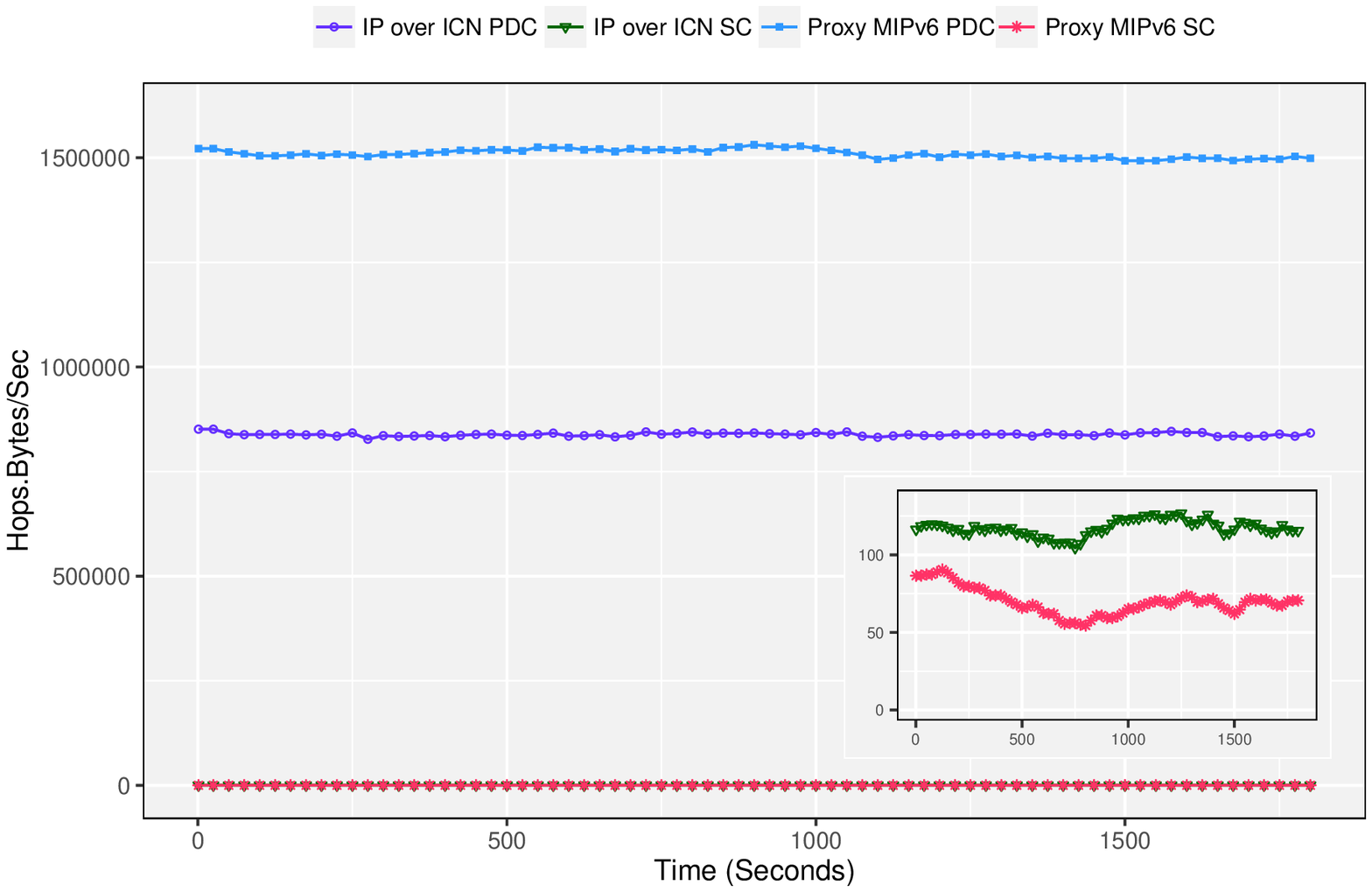}
			\caption{Single Speed (70 miles/h).}
			\label{fig:constav}
		\end{subfigure}
		~
		\begin{subfigure}[b]{0.5\textwidth}
			\includegraphics[width=\columnwidth,height=2.8in]{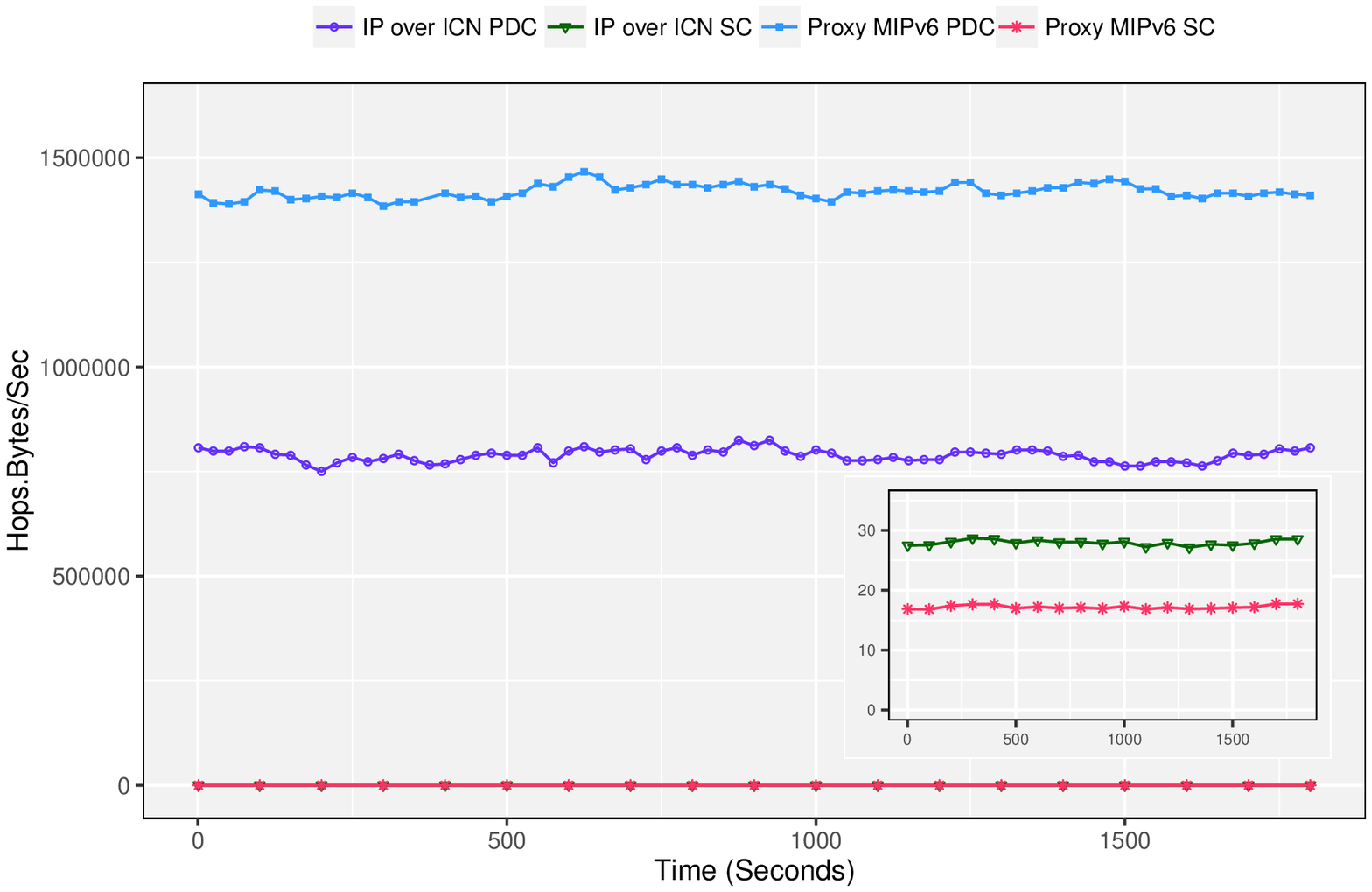}
			\caption{Random Speed (3 - 70 miles/h).}
			\label{fig:randav}		  
		\end{subfigure}
		\caption{Simulation of 50 MN's with average packet delivery cost (PDC) and signaling cost (SC) using a Geometric Random Topology of 100 nodes.}
	\end{figure*}
\end{center}
\vspace{-5mm}
\section{Mobility Management Simulation and Cost Evaluation}
To evaluate the performance of the proposed IP-over-ICN mobility solution and show the significance of the established analytical model, a discrete time event simulation of both Proxy Mobile IP and IP-over-ICN has been conducted in R. The built simulation environment has been used to investigate the mobility costs (mainly mobility signaling, packet delivery and handover latency costs) with different scenarios and compare the results with those of our analytic model. Random geometric networks have been used to represent network topologies in our simulation to ensure spatial homogeneity of the positions of the MAGs and NAPs. Various network topology sizes with a different number of nodes (MAG's and NAPs) have been simulated with varying node degree. Both MAG's and NAPs have been simulated using a circular coverage area with a radius of 500 m. The same central node was used to represent the LMA and TM/RV in all cases to ensure valid cost comparisons. A random walk mobility model has been used to capture user mobility with various speed values ranging from pedestrians moving at a rate of 3 miles/hour to vehicles travelling at 70 miles/hour. Initial user locations are randomly distributed using a uniform random distribution. In our traffic model, we assume that all the users in the network exchange video data with an arrival rate of 1 Mbps following a Poisson distribution. Every simulation experiment was run for 1800 seconds and repeated 20 times with results collated after reaching a steady state.

\subsection{Validating the Analytical Model Through Simulation}
The first simulation results use the topology example in Fig. \ref{top2} to compare the performance with those obtained from the cost analysis functions and mobility model in the previous section. Our modelling results where calculated as follows. Assuming a vehicle moving at a constant velocity of 70 mile/h through a circular coverage area with a radius of 500 m, this would result in the vehicle spending 25.12 second in every cell it traverses at a mobility rate $(\mu)$ of 0.039  \;1/s. Therefore, applying the derived cost functions for PMIPv6 and IP-over-ICN in equations (\ref{emc7}), (\ref{emc8}), (\ref{emc12}) and (\ref{emc15}) to the network model example in Fig. \ref{top2}; and utilizing the Markov mobility model in Fig. \ref{fig:mc} while substituting the variables with the values in Table \ref{top2}; yields in the following results. $\Upsilon$ = 21.624 hops.Bytes$/$s, $\Lambda$ = 809176 hops.Bytes$/$s, $\Upsilon^\prime$ = 43.758 hops.Bytes$/$s and $\Lambda^{\prime}$ = 218400 hops.Bytes$/$s. To compare the results with those that would be obtained from a simulation that uses the random walk mobility model, a single MN was simulated to move randomly with speed of 70 mile/h. Both the MN initial location and traversed paths were selected randomly from a uniform distribution. Fig. \ref{fig:simvsmodel1} shows the accumulative simulation and modelling results for the example topology in Fig. \ref{top2} over 1800 seconds. The results show that the modelling results fall withing the 95\% confidence intervals of the simulation results showing that there is a high degree of certainty that the two methods agree. Fig. \ref{fig:simvsmodel1} also shows both the total packet delivery and signaling costs for Proxy MIPv6 and IP-over-ICN, and it is clear from the results that Proxy MIPv6 imposes a higher packet delivery cost of more than three times that of IP-over-ICN reaching about $15\times 10^8$ Hops.Bytes due to the longer traffic paths imposed by using a central LMA. Also from Fig. \ref{fig:simvsmodel1} it can be seen that IP-over-ICN imposes higher signaling cost than Proxy MIPv6 reaching approximately $9\times 10^4$ Hops.Bytes compared to about $4\times 10^4$ Hops.Bytes incurred by Proxy MIPv6. But despite the signaling cost results, the high difference in magnitude of packet delivery cost between Proxy MIPv6 and IP-over-ICN indicates that IP-over-ICN highly outperforms PMIPv6 in the total cost. 

\subsection{Mobile Node Speed Variation}
\begin{center}
 	\begin{figure*}
 		\begin{subfigure}[b]{0.5\textwidth}
 			\includegraphics[width=\columnwidth]{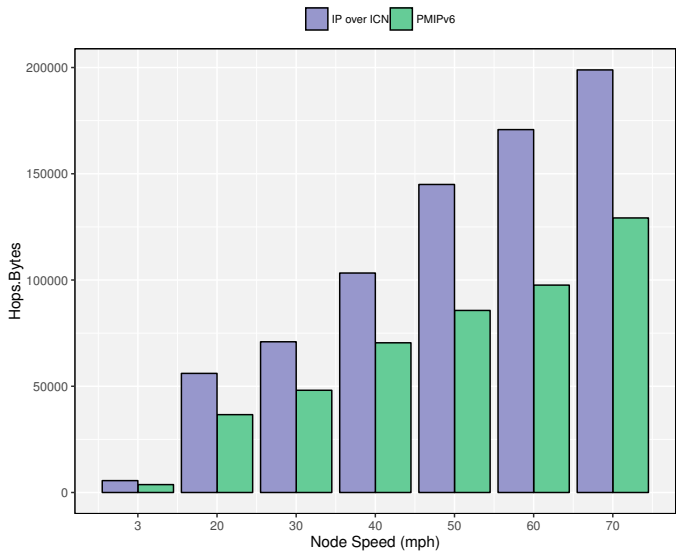}
 			\caption{ Mobility Signaling Cost with Different Nodes Speed. } 
 			\label{diffspeed}
 		\end{subfigure}
 		~
 		\begin{subfigure}[b]{0.5\textwidth}
 			\includegraphics[width=\columnwidth]{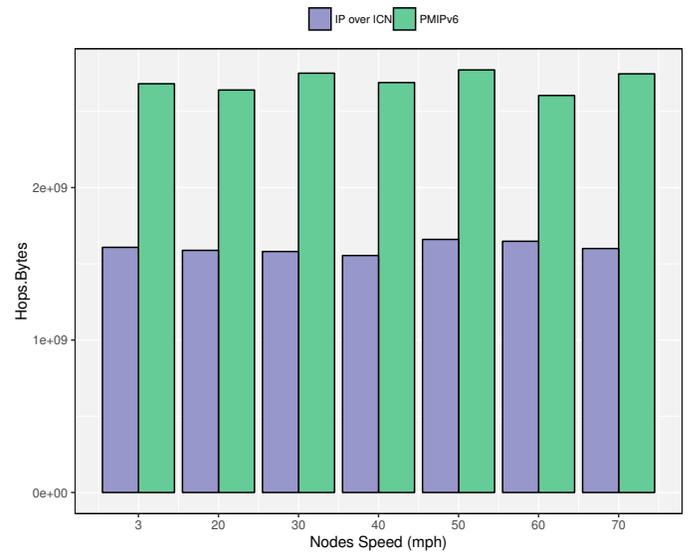}
 			\caption{ Mobility Packet Delivery Cost with Different Nodes Speed. } 
 			\label{diffspeed2}
 		\end{subfigure}
 		\caption{Simulation of 50 MN's with total packet delivery cost (PDC) and signaling cost (SC) using a Geometric Random Topology of 100 nodes.}
 	\end{figure*}
 \end{center}

The second set of results use random geometric networks of 100 nodes with average connection degree of 4 neighbors (between 1 and 8 neighbors for every NAP/MAG in the network). 50 MNs where simulated to roam freely and randomly within the network domain. Various node speeds have been used in this experiment ranging between pedestrian speeds of 3 miles/h and highway speeds of 70 miles/hour. The MN initial locations, traversed paths and speeds, were all selected randomly from uniform distributions. Fig. \ref{fig:constav} shows the average packet delivery and signaling costs at 70 miles/h for both Proxy MIPv6 and IP-over-ICN. According to this figure, Proxy MIPv6 shows approximately double the packet delivery cost imposed by IP-over-ICN due to the central traffic anchoring. Although IP-over-ICN shows higher signaling costs, the high difference in figures between packet delivery cost and signaling cost implies that IP-over-ICN requires half the total cost of Proxy MIPv6 in order to provide network enabled mobility support. Another simulation run is shown in Fig. \ref{fig:randav} using random MN speeds ranging from 3 to 70 miles/h. The figure clearly shows that the same difference in performance is observed between IP-over-ICN and Proxy MIPv6 in terms of packet delivery and signaling costs respectively although random MN speeds have been simulated. Figs \ref{diffspeed} and \ref{diffspeed2} show the total incurred mobility signaling and packet delivery costs respectively when different MN speeds are simulated individually. From the results it can be seen that mobility signaling cost has a positive relation with MN speed ranging from only 3720 and 5593 Hops.Bytes for Proxy MIPv6 and IP-over-ICN respectively with MN speed of 3 miles/h to about 1.3 x $10^6$ and 2 x $10^6$ Hops.Bytes for Proxy MIPv6 and IP-over-ICN respectively with MN speed of 70 miles/h. On the other hand, the packet delivery cost is not influenced by any speed changes as seen in Fig. \ref{diffspeed2}.

\subsection{Network Topology Size Variation}
\begin{figure}[t]
 	\centering
 	\includegraphics[width=\columnwidth]{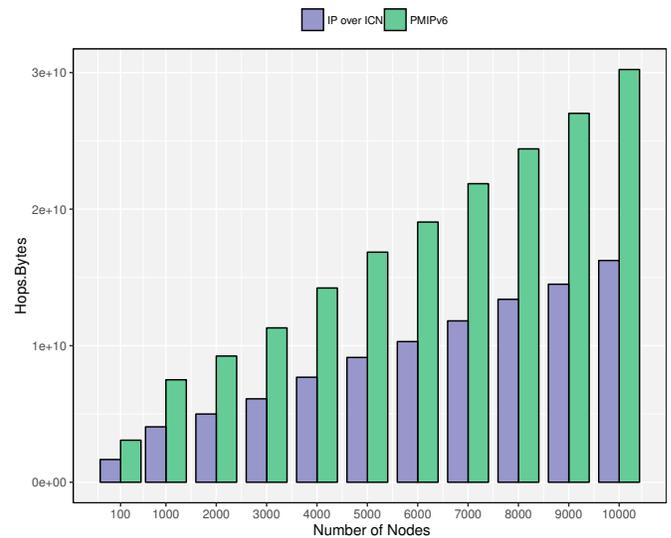}
 	\caption{ Total Cost with Different Network Topology Sizes. } 
 	\label{diffsize}
 \end{figure}

The third simulation results are conducted using different network sizes to show how network dimensions can effect the total cost. Random geometric networks ranging from 100 up to 10000 nodes have been simulated with average connection degree of 4 neighbors for every NAP/MAG in the network. 50 MNs where simulated to roam freely and randomly within the network domain with speed of 70 miles/h. Fig. \ref{diffsize} shows the total cost (packet delivery + signaling) for both Proxy MIPv6 and IP-over-ICN for each of the simulated topology sizes. It can be seen from the trends that IP-over-ICN always outperforms Proxy MIPv6 in terms of the total cost required to support network-based mobility management with an improvement factor of at least 1.8 due to the sub-optimal triangular routing mechanism of Proxy MIPv6.
\vspace{-5mm}
\subsection{Handover Latency}
\begin{figure}[t]
 	\centering
 	\includegraphics[width=\columnwidth,height=2.8in]{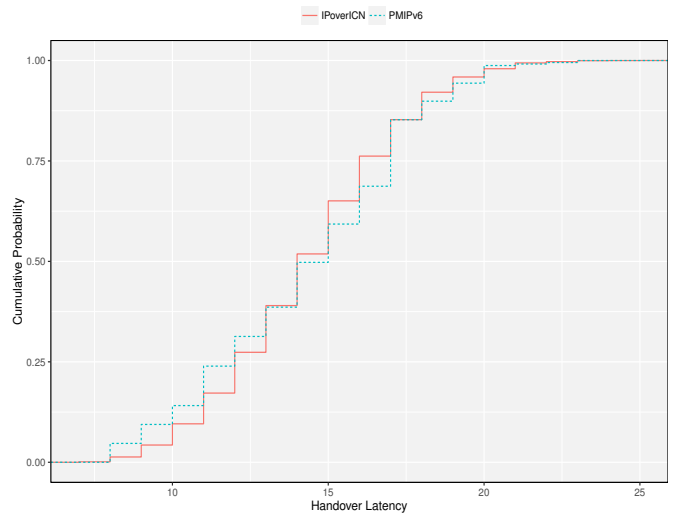}
 	\caption{Empirical Distribution Function of Handover Latency.}
 	\label{eCDF}
 \end{figure}

The final simulation experiment examines the handover latency in both IP-over-ICN and PMIPv6 networks using the same network topology of Section 6.2 with 100 MNs roaming freely and randomly within the network domain at 70 miles/hour. Fig. \ref{eCDF} shows an empirical cumulative distribution function of the handover latency in both investigated domains. From this figure it can be seen that IP-over-ICN and PMIPv6 networks have highly convergent distributions that are nearly identical at the upper range of handover latencies between 20-25 unit time. This is due to the high similarity in the number of signalling messages and processes required to facilitate handover in both domains with minor differences in the traversed paths as outlined in section 5.3. This clearly illustrates that IP-over-ICN offers no extra cost in handover latency while earlier results show significant savings on the data plane traffic.
 
 	\vspace{-5mm}
\section{Related work}
  Proxy MIPv6~\cite{gundavelli2008b} has been adopted by the IETF to support network-based mobility in wireless networks by utilizing the LMA as a centralized mobility management entity on both the data and control plane. On the other hand, 3GPP specifies the General packet radio service (GPRS) Tunnelling Protocol (GTP)~\cite{lucent2009lte} to support mobility in cellular networks by anchoring user data plane traffic at the serving gateway (S-GW) and control plane traffic at the MME. GTP is an important IP/UDP based protocol used to encapsulate user data when passing through core network using GTP-U and also carries bearer specific signaling traffic between various core network entities using GTP-C. Also in efforts to significantly improve handover between heterogeneous network technologies, IEEE standards association has developed 802.21~\cite{taniuchi2009ieee} that defines a media-independent handover (MIH) framework. The standard defines the tools required to exchange information, events, and commands to facilitate handover initiation and handover preparation. A large number of efforts have focused on amendments, improvements and cost evaluation of the standards mentioned previously, we summarize the most significant of them below.
 
 Distributed Mobility Management (DMM) efforts \cite{liu2015distributed} \cite{bernardos2013pmipv6} try to solve Proxy MIPv6 drawbacks by evolving towards a flatter architecture using distributed anchoring, thereby providing a more efficient way to handle mobile traffic. In these approaches, although the LMA functionality is distributed into the network edges, they still perform traffic tunnelling and anchoring in a localized manner which does not eliminate the traffic overhead imposed to support mobility.
 
 IEEE 802.21 Media Independent Handover (MIH) functionality assisted Proxy MIPv6 solutions such as \cite{kim2013ieee}, aim at reducing handover latency and signaling cost in heterogeneous wireless networks. The base station with MIH functionality performs handover on behalf of the MN. The analytical evaluation shows that the proposed mechanism can outperform the existing mechanism in terms of handover latency and total number of over the air signaling messages. But despite that, the sub-optimal core routing problem remains unsolved.
 
 Path-based forwarding architectures such as Software Defined Networks (SDN), bring new possibilities to improve the mobility management with lower traffic cost, better scalability and faster handover. Today, the most known approach is testing mobile flow entries against matching rule fields and finding a correct output action through every OpenFlow switch along the path, which has high costs in mobile flow management. Most of the proposed SDN architectures in wireless networks cannot be directly applied to large-scale networks due to this reason \cite{dai2016core}. OpenFlow-enabled proxy mobile IPv6 (OF-PMIPv6) is proposed in \cite{raza2016leveraging} where the control path is separated from the data path by performing the mobility control at the controller, whereas the data path remains direct between the MAG and the LMA in an IP tunnel form. This method achieves improved handover latency over conventional Proxy MIPv6, while the data plane anchor problem is persistent. Other SDN efforts such as \cite{dong2015rule} propose rule caching mechanisms to tackle the limited rule space problem in existing SDN devices. Such approaches propose to support completely flat mobility architectures, but as a drawback, they incur additional processing complexity to manage the proposed caching mechanisms.

\vspace{-3.6mm}
\section{Conclusion}
Efficient mobility management solutions are essential to accommodate the immense growth of mobile networks, users and generated traffic. In this paper we introduced a novel, anchor-free, mobility management solution that utilises a revolutionary path-based forwarding substrate to enable direct communication between the source and destination. We evaluated the cost of our solution through analytical modelling and simulations; and, compared it with the conventional PMIPv6. Evaluation results have shown that the delivery cost of our solution is approximately half that incurred by the PMIPv6 counterpart; for similar or (in some cases) reduced end-to-end latency. Consequently, we have shown that significant resource saving can be achieved using our proposed solution.

By introducing the anchor point, PMIPv6 clearly violates network end-to-end transparency, and also introduces a network state (not flow-based, but device based), which is considered a drawback for processing, security as well as failure perspectives. Strictly speaking, IP-over-ICN still violates the transparency but at a much better point of the system, namely at the attachment points of both communication parties. This paper demonstrates that this is an improved point for the violation, as it allows optimal delivery paths \emph{i.e.} the same path that would be used if mobility had not occurred. This paper has used an IP-over-ICN solution as an embodiment to facilitate the proposed anochor-free mobility solution. However, the proposed mobility solution can be facilitated by any forwarding architecture that purely relies on path information stored in the forwarded packet for the end-to-end delivery; in this case, mobility simply results in partial recomputation of the path, with the opportunity to deliver the data over an optimal path after every handover operation. 
\vspace{-4mm}
\bibliographystyle{IEEEtran}
%
%
%
\bibliographystyle{IEEEtran}
\bibliography{ref}
%




\end{document}